\documentclass[preprint,nofootinbib,floatfix]{revtex4}

\usepackage[british]{babel}
\usepackage{amsbsy}
\usepackage{amssymb}
\usepackage[sumlimits,tbtags]{amsmath}
\usepackage{graphicx}
\usepackage{subcaption}
\usepackage{xcolor}
\usepackage[active]{srcltx}
\usepackage{pdfsync}
\usepackage{xfrac}

\usepackage{mathtools}
\usepackage{ulem}

\usepackage{lipsum}

\usepackage[hyperfootnotes = false]{hyperref}
\usepackage{cleveref}
\hypersetup{
 colorlinks,
 linkcolor=blue,
 citecolor=red,
 urlcolor=blue, 
 }
 
 \usepackage{physics}

\usepackage{color}

\usepackage{cancel}

\newcommand{\la}{{\langle}}
\newcommand{\ra}{{\rangle}}

\def\b{{\rm b}}

\def\e{{\rm e}}

\def\l{{\rm l}}
\def\n{{\rm n}}
\def\p{{\rm p}}
\def\q{{\rm q}}
\def\s{{\rm s}}
\def\x{{\mathrm{x}}}

\newcommand{\A}{{\cal A}}
\newcommand{\B}{{\cal B}}

\newcommand{\E}{{\cal E}}

\def\O{{\cal O}}

\def\tn{{\tilde{n}}}
\def\teps{{\tilde{\varepsilon}}}
\def\tu{{\tilde{u}}}

\def\tT{{\tilde{T}}}
\def\ts{{\tilde{s}}}
\def\tmu{{\tilde{\mu}}}
\def\tp{{\tilde{p}}}
\def\tperp{{\tilde{\perp}}}

\def\tY{{\tilde{Y}}}
\def\tnu{{\tilde{\nu}}}
\def\tbeta{{\tilde{\beta}}}

\def\eps{{\epsilon}}
\def\veps{{\varepsilon}}
\def\om{{\omega}}

\newcommand{\wh}{\hat{\omega}}
\newcommand{\whd}{\hat{\omega}_{\Delta}}

\newcommand{\comment}[1]{}

\newcommand{\change}[1]{#1}

\begin{document}

\title{Formulating bulk viscosity for neutron star simulations}

\author{T. Celora$^1$, I. Hawke$^1$, P.C. Hammond$^1$, N. Andersson$^1$, G. L. Comer$^2$}

\affiliation{
$^1$ Mathematical Sciences and STAG Research Centre, University of Southampton,
Southampton SO17 1BJ, United Kingdom\\
$^2$ Department of Physics, Saint Louis University,
St Louis, MO 63156-0907, USA }

\begin{abstract}
In order to extract the precise physical information encoded in the gravitational and electromagnetic signals from powerful neutron-star merger events, we need to include as much of the relevant physics as possible in our numerical simulations. This presents a severe challenge, given that many of the involved parameters are poorly constrained.
In this paper we focus on the role of nuclear reactions.
Combining a theoretical discussion with an analysis connecting to  state-of-the-art simulations, we outline multiple arguments that lead to a reactive system being described in terms of a bulk viscosity. The results demonstrate that in order to properly account for nuclear reactions, future simulations must be able to handle different regimes where rather different assumptions/approximations are appropriate. 
We also touch upon the link to models based on the large-eddy-strategy required to capture turbulence. 

\end{abstract}

\maketitle

\section{Introduction}

Neutron star mergers involve violent dynamics of high-density matter in the live spacetime of general relativity. These events provide an opportunity to explore many extremes of physics, from the state of matter beyond the nuclear saturation density (the elusive equation of state \cite{Oertel+EoSrev,BurgioFantinaEoSRev,VidanaEoSRev,Lattimer2021})  to the formation of a long-lived merger remnant (likely a black hole \cite{ShibataHotokezakaRev2019,BernuzziRemnant,BaiottiRezzollaRev}). Matter outflows and the associated rapid nuclear reactions determine the (hopefully) observable kilonova signature \cite{MetzgerKilo} and the twisting of the stars' magnetic field may help collimate an emerging jet and hence explain observed short gamma-ray bursts \cite{Paczynski,CiolfiRev}. The spectacular GW170817 event \cite{LIGOScientific:2017vwq,LIGOScientific:2017ync} demonstrated the range of exciting features \cite{Bauswein_2017,Rezzolla_2018,Margalit_2017,RuizPRD18,ShibataPRD19}. With more sensitive instruments coming on-stream in the future (like the Einstein Telescope \cite{ETScienceCase} and the Cosmic Explorer \cite{CEbeyondLIGO,CEScienceCase} for gravitational waves) higher precision observations are anticipated and one may hope to extract more detailed information about the involved physics. This is obviously an exciting prospect, but it prompts us to consider two important issues. First, from the observational point of view, can we ascertain that the physics we hope to explore is cleanly associated with a particular observational feature? Second, from the modelling perspective, are we sure that these features are faithfully represented by our theoretical models? Specifically, how do we know that we ``get the physics right'' in our numerical simulations? Given the complexity of the problem---involving the dynamics and thermodynamics of out-of-equilibrium matter at extreme densities and temperatures---this is a very challenging question. 

In this paper we take modest steps towards a partial answer of the modelling question. We consider the role of bulk viscosity associated with nuclear reactions---which may, or may not, leave an observable imprint on (say) the gravitational-wave signal \cite{Most2021BV}---and ask how this mechanism can be implemented in nonlinear simulations. In particular, we highlight the formal aspects of the problem and establish how the inevitable ``limitations'' of a numerical simulation (in terms of resolution) enter the discussion. The aim is to establish to what extent simulations based on an effective bulk viscosity are viable and (perhaps more importantly) when they are not. This understanding will be crucial for future numerical implementations.  

\section{Stating the problem}

The underlying physical model of a neutron star merger is expected to be a system of multiple interacting ``fluids'' of different charged particle species, coupled to an electromagnetic field and radiation through, for example, neutrinos, all evolving on a dynamical relativistic spacetime \cite{livrev}. The complexity of this model makes it impractical for use in either theoretical calculations or numerical simulations. Instead, simplifications must be made, with heuristic arguments needed to justify each assumption that is introduced.

To illustrate the key argument with a simple toy model, consider the problem of heat propagation. When the underlying model is required to be \emph{causal}, the starting point is often taken to be the Cattaneo equation (here given in $1+1$ dimensions, see \cite{EIT})
\begin{subequations}
\label{eq:cattaneo}
\begin{align}
    \label{eq:cattaneo_t}
    \pdv{T}{t} + \pdv{q}{x} &= 0, \\
    \label{eq:cattaneo_q}
    \pdv{q}{t} &= -\tau^{-1} \left[ \kappa \pdv{T}{x} + q \right].
\end{align}
\end{subequations}
In the ``equilibrium limit'', when the relaxation time $\tau \to 0$, we recover Fourier's law 
\begin{equation}
    q = -\kappa \pdv{T}{x}
\end{equation} 
relating the heat flux $q$ to the temperature $T$,  from~\cref{eq:cattaneo_q}. This then  reduces~\cref{eq:cattaneo_t} to the familiar heat equation. While the underlying model is hyperbolic, the fast relaxation limit is parabolic and hence not causal.

Specifically, working out the characteristic velocities in the problem (through the standard dispersion relation) one finds that the Cattaneo equation~\cref{eq:cattaneo} is causal with finite propagation speeds bounded by $\pm (\kappa / \tau)^{1/2}$. At the same time, there is a critical wavenumber $\gtrsim (\kappa \tau)^{-1/2}$ below which the behaviour is purely parabolic and the solution is diffused away. An illustration of the transition from second sound to diffusion can be found in Figure 16 of~\cite{livrev}.

From a theory point of view it would be natural to argue that we should base our models on the Cattaneo formulation, but from a numerical perspective this may be problematic. We would need to resolve the (presumably fast) relaxation towards equilibrium and this may not be possible/practical. In this sense, the parabolic prescription may be preferable. 

A heuristic argument for using the parabolic heat equation within a relativistic model---for which causality would be a prerequisite---would be as follows. We assume that on the length scales $L$ relevant for our model we have $\tau \sim L / c$, where $c$ is the speed of light. By causality there can be no (propagating!) scales of physical relevance faster than $c$, hence with timescales smaller than $\tau$ or (equivalently) frequency scales larger than $\tau^{-1}$. Therefore the only relevant behaviour for heat propagation is the purely parabolic case where heat fluctuations are rapidly damped, which is well modelled by the standard heat equation.
It is possible to use the internal consistency of the underlying model to check when this heuristic argument is valid. For example, to be consistent with causality the dispersion relation at low frequencies requires that $\kappa \le \tau c^2 \sim c L$.

The key point here is the existence of a single scale (length, time, or frequency) at which some physical effect acts or changes type. This issue is often considered for turbulence (for example, are the length scales probed sufficient to trigger the magneto-rotational instability \cite{Duez+MRI,SiegelMRI,KiuchiMRI,GuiletMRI}), for reactions (are physical regions of parameter space probed so that out-of-equilibrium physics, such as a  bulk viscosity has to be accounted for \cite{PRLAflordBulkMergers}), and for radiation (are neutrinos propagating or trapped, see e.g.\ \cite{PeteThermal}). When this key scale is outside that which can be included in the model then the physical effect is either ignored (by using a purely ideal model, or by assuming instantaneous relaxation to an equilibrium) or modelled (by approximating the additional physics through a closure term, such as an effective equation of state, or a large-eddy approach \cite{PeteThermal,Giacomazzo,radice1,carrasco}).

The fundamental issue regards how the physics, and hence the model, should behave when key scales cross or overlap at different parts of the required parameter space. This is particularly relevant for nonlinear numerical simulations, where the discretization length introduces a scale (or multiple scales with uneven grids or mesh refinement), and nonlinearity can lead to the relevant physical scales varying over many orders of magnitude. The interaction between the different scales makes  heuristic simplifications dubious.

With these issues in mind we will study, analytically and numerically,  issues relating to bulk viscosity in reactive fluids. Here the underlying model involves nuclear reactions, specifically the direct and modified Urca reactions (although the analytical calculations presented apply more generally). These microphysical reactions can, in some regimes, be approximated as a (resonant) bulk viscosity. However, the timescales on which the reactions take place strongly depend on (for example) the temperature and may as a result be close to those that can be captured in numerical calculations. The timescales may also interact with, for example, large eddy closure terms required for turbulent regions (cf. \cref{sec:LES}).

Before we proceed, it may be worth commenting on the layout of the paper. The problem we consider requires us to bring together a number of different aspects, some theoretical (relating to the involved physics) while others relate to the numerical implementation (the impact of a finite resolution etcetera). In order to avoid confusion, many of the details are provided in (unusually lengthy) appendices. The aim is to keep the presentation as ``linear'' as possible, allowing the reader to appreciate the main points and the spirit of the proposed formulation. A more detailed reading would involve a closer study of the appendices, which back up the arguments in the main text.

\section{Hydrodynamics of a reactive system}\label{sec:hydrodynamics}

We want to consider the hydrodynamics of an isotropic reactive system consisting of comoving neutrons, electrons and protons, the simplest meaningful matter composition for a neutron star core.\footnote{This may seem somewhat reductionist, given that the high density region is likely to bring other matter constituents into play, but our main interest is to establish the principles involved. If richer matter models are required then the extension of our discussion  will be conceptually straightforward.} Further, because we assume charge neutrality, there are only two independent number densities (or, equivalently, one density and one species fraction). These are conveniently taken as $n$, representing the baryon number density, and $Y_\e = {n_\e}/{n}$ the electron (lepton) fraction. The proton fraction then follows as $Y_\p = n_\p/n = Y_\e$ while the neutron fraction is given by $Y_\n = n_\n/n = 1 - Y_\e $. Because we assume isotropy, the stress-energy tensor takes the perfect fluid form 
\begin{equation}\label{eq:SEMtensor}
	 T^{ab} = \veps  u^a u^b +  p (g^{ab} + u^a u^b) \;,
\end{equation}
where $a$ and $b$ are spacetime indices and the four-velocity $u^a$ is defined uniquely as the one associated with the flow of all particle species. The pressure and energy are identified with the corresponding thermodynamical quantities. As we will not impose that the system is in chemical equilibrium (we allow reactions) we assume the equation of state to involve three parameters. That is, the pressure follows from $p = p (n,\veps, Y_\e)$. This relation is assumed to be provided in tabulated form (suitable for a numerical simulation, see \cref{app:compOSE} and \cite{compOSE}). 
The energy-momentum conservation laws can be written as usual as
\begin{subequations}\label{eqs:Euler}
\begin{align}
    u^b\nabla_b \veps &= - (p + \veps) \theta \;,\\
    (p+ \veps) a_b &= - \perp_b^c \nabla_c p \;,
\end{align}
\end{subequations}
where $a^b = u^a\nabla_a u^b$,  $\theta = \nabla_a u^a$ is the fluid expansion  and $\perp^{ab} = g^{ab} + u^a u^b$. We also need equations for the particle number densities. These are
\begin{equation}\label{eq:baryonCont}
	u^a \nabla_a n + n \theta = 0 
\end{equation}
for the baryon number density (which is obviously conserved), and
\begin{equation}\label{eq:ElectronFractionEq}
	u^a \nabla_a Y_\e = \frac{\Gamma_\e}{n}
\end{equation}
where the rate $\Gamma_\e$ is generally non-vanishing as we are considering a reactive system. Once the reaction rate is provided by the microphysics (and tabulated as a function of the other variables), these equations constitute a closed system. 

From the perspective of  thermodynamics, it is natural to introduce the affinity $\beta = \mu_\n - \mu_\p - \mu_\e$ as it quantifies how far the system is out of cold beta equilibrium. This has the advantage that we can work within the so-called Fermi surface approximation \cite{AlfordHarris18} and express relevant quantities, like the reaction rates, as expansions for small values of $\beta$ (with the coefficients in the expansion evaluated at equilibrium, $\beta=0$). In the following, and notably in the illustrations we provide, we assume that this strategy is appropriate. Pragmatically, this makes sense as we are only aiming to establish a proof of principle and these assumptions allow us to work out all required parameters for the model from a standard tabulated equation of state. However, it is important to keep in mind that the assumptions will not be appropriate for much of the parameter space (temperature and density) explored in the binary neutron star merger/post-merger phase, and they completely exclude any neutrino effects. At finite temperatures, the true notion of beta equilibrium is more complex, and may require the addition of an isospin chemical potential in the definition of $\beta$ (see \cite{AlfordHarris18,AlfordHarrisDamping18,Alford_Zhang21,AlfordHarutyunuanSedrakian19,AlfordHarutyunuanSedrakian21,PeteThermal}). A complete model should account for the correct notion of equilibrium, but  this will require the equation of state table to be extended to include all necessary information. As such data is not yet available for simulations, we are (pragmatically) doing the best we can given the information at hand.

As we will see in \cref{sec:thermodynamics}, the affinity is (thermodynamically) conjugate to $Y_\e$, meaning that either of the two variables can be used ``equivalently'' in the discussion. This is important because state-of-the-art simulations tend to involve $Y_\e$ while the theory is somewhat more transparent when expressed in terms of $\beta$. In the following we will develop the model in terms of $\beta$, but the analogous arguments for $Y_\e$ can be found in \cref{app:ElectronFraction}. The evolution equation for $\beta$ is easily obtained by considering it as a function of $(\veps,n,Y_\e)$. We arrive at 
\begin{equation}\label{eq:BetaEqGeneral}
	u^a \nabla_a \beta = \left(\frac{\partial \beta}{\partial Y_\e}\right)_{n,\veps}\frac{\Gamma_\e}{n} - n\B\theta
\end{equation}
with 
\begin{equation}\label{eq:DefB}
    \B = \left(\frac{\partial \beta}{\partial n}\right)_{\veps,Y_\e} + \frac{p+\veps}{n}\left(\frac{\partial \beta}{\partial \veps}\right)_{n,Y_\e} .
\end{equation}
We again see that the system of equations is closed once the reaction rate (as well as the relevant thermodynamical coefficients) is provided. 

A useful simplification occurs when the system is sub-thermal, when ${\beta}/{T}\ll 1$. Then we can expand the rate with respect to chemical equilibrium $\beta = 0$ to write it as\footnote{There is a sign convention here, and we are following \cite{AlfordBulk10}. The logic is, if $\beta > 0 $ then $\mu_\n >\mu_\e + \mu_\p$ and neutron decay is favoured (over electron-capture) as this will release energy. Therefore, we want the electron rate to be positive when $\beta$ is positive and vice versa. The sign of $\gamma$ should then be negative.} $\Gamma_\e = -  \gamma \beta$. The evolution equation for the affinity $\beta$ then simplifies to
\begin{equation}\label{eq:BetaEquationSubthermal}
	u^a \nabla_a \beta = - \A \beta - n\B \theta
\end{equation}
where we introduced the new coefficient (with units of inverse time)
\begin{equation}\label{eq:Adef}
    \A = \frac{\gamma}{n}\left(\frac{\partial \beta}{\partial Y_\e}\right)_{n,\veps} \;.
\end{equation}
The information encoded in the reaction rate $\Gamma_\e$ is now ``stored'' in $\gamma$. We can then make progress and compute $\gamma$ from the equation of state tables provided in the compOSE database \cite{compOSE} \underline{only as long as} we ignore finite temperature effects \cite{AlfordHarris18,AlfordHarrisDamping18,Alford_Zhang21,AlfordHarutyunuanSedrakian19,AlfordHarutyunuanSedrakian21}. 
While the coefficient $\B$ can be introduced without reference to an expansion around equilibrium, this is not the case for $\A$. In the sub-thermal limit we  retain only terms linear in $\beta$, so that $\A$ must be evaluated at $\beta = 0$. 

For completeness, let us also comment on the entropy density, viewed as a function of $(\veps, n, Y_\e)$. Using the equations of motion for these quantities (as well as the generalized Gibbs relation provided in \cref{sec:thermodynamics} below) we arrive at 
\begin{equation}
  T \nabla_a \left( su^a \right) =  \beta\Gamma_\e\;.
\end{equation}
As long as $\beta$ has the same sign as $\Gamma_\e$ the entropy  increases. This is guaranteed to be the case in the sub-thermal limit if $\gamma < 0$. Note that this assumes that a negligible amount of energy is deposited in neutrinos by the reactions, which will be a poor approximation at high temperatures. This important caveat will quantitatively affect our results without changing the qualitative conclusions.

\section{Thermodynamics of a reactive system}\label{sec:thermodynamics}

Having discussed the hydrodynamics, let us turn to the associated thermodynamics. Because the underlying model is required to be causal, and hence based on Cattaneo-type laws, it is natural to set the discussion within the Extended Irreversible Thermodynamics (EIT) framework \cite{livrev}. We here provide a streamlined discussion, and refer to the monograph \cite{EIT} and references therein (see also \cite{GavassinoBulk} for a recent analysis focused specifically on bulk viscosity).
The first step is to \underline{assume} that the Gibbs relation takes the usual form---noting that the various quantities may not be in thermodynamical equilibrium
\begin{equation}\label{eq:OutEqGibbs}
    p+\veps = \sum_{\x = \n,\p,\s,\e} n_\x \mu_\x = n\mu_\n - n_\e \beta + T s
\end{equation}
and
\begin{equation}\label{eq:pressuredifferential}
    dp =\sum_{\x = \n,\p,\s,\e} n_\x d\mu_\x  = n d\mu_\n - n_\e d\beta + s dT .
\end{equation}
As we will be working at the fluid level with either $(n,\veps,Y_\e)$ or $(n,\veps,\beta)$, we can use the entropy as a thermodynamical potential (see \cref{app:subsec:Yethermo} for an equivalent discussion and results that make explicit use of a notion of equilibrium electron fraction). This is convenient because if we also assume that the system is close to thermodynamical equilibrium, we can  expand the entropy as 
\begin{equation}
    s = s^\mathrm{eq}(n,\veps) + \frac{1}{2}s_2(n,\veps) \beta^2 \;, \quad \text{where } s_2 = \left(\frac{\partial^2s}{\partial\beta^2}\Big|_{\beta = 0}\right)_{n,\veps} .
\end{equation}
From this we can compute the out-of-equilibrium expansion of the thermodynamical variables. Linearizing in deviations from equilibrium, and assuming that the equation of state is expressed in terms of $Y_\e$ rather than $\beta$, we obtain
\begin{subequations}\label{eq:expansionsWithBeta}
\begin{align}
    T &= T^\mathrm{eq} \left[ 1 -n \left(\frac{\partial \beta}{\partial \veps}\right)_{n,Y_\e} \left(\frac{\partial \beta}{\partial Y_\e}\right)_{n,\veps}^{-1} \beta\right]  + \O(\beta^2) \;,\\
    \mu_\n &= \mu^\mathrm{eq} - \left[ n \mu^\mathrm{eq} \left(\frac{\partial \beta}{\partial \veps}\right)_{n,Y_\e} + n  \left(\frac{\partial \beta}{\partial n}\right)_{\veps,Y_\e}- Y_\e \right]\left(\frac{\partial \beta}{\partial Y_\e}\right)_{n,\veps}^{-1} \beta + \O(\beta^2) \;, \\
    s&= s^\mathrm{eq} + \frac{1}{2} \frac{n}{T^\mathrm{eq}}\left(\frac{\partial \beta}{\partial Y_\e}\right)_{n,\veps}^{-1}\beta^2 \;.
\end{align}
\end{subequations}
Note that the thermodynamical requirement on the entropy reaching a maximum at equilibrium, namely $s_2 < 0$, implies that ${\partial \beta}/{\partial Y_\e}$ must be negative. Recalling \cref{eq:BetaEquationSubthermal,eq:Adef}, and the fact that the restorative term $\gamma<0$, we see that $\A >0$ and therefore plays the role of an (inverse) relaxation rate.

Now that we have  expansions (in $\beta$) of the thermodynamical variables, we can use the Gibbs relation to work out the pressure. To linear order in the deviation from equilibrium we then have 
\begin{equation}
    p = p(n,\veps,\beta) = p^\mathrm{eq} + p_1 \beta \;,\quad \text{where } p^\mathrm{eq} = p(n,\veps,\beta = 0) \;,\; \text{and } p_1 = \left(\frac{\partial p}{\partial \beta}\Big|_{\beta=0}\right)_{n,\veps} \;,
\end{equation}
or, explicitly, 
\begin{align}
    p^\mathrm{eq} = -\veps + n\mu^\mathrm{eq} + T^\mathrm{eq} s^\mathrm{eq} \;,
\end{align}
and 
\begin{equation}\label{eq:p1WithBeta}
    p_1 = - n^2 \left(\frac{\partial \beta}{\partial Y_\e}\right)_{n,\veps}^{-1} \left[\frac{p^\mathrm{eq} + \veps}{n}\left(\frac{\partial \beta}{\partial \veps}\right)_{n,Y_\e} + \left(\frac{\partial \beta}{\partial n}\right)_{\veps,Y_\e} \right] \;.
\end{equation}
In essence, the thermodynamical expansion provides us with an expression for the out-of-equilibrium contribution to the pressure, which would naturally be interpreted as a bulk viscosity. We identify
\begin{equation}\label{eq:Pi_t}
    \Pi_t = \left(\frac{\partial p}{\partial \beta}\Big|_{\beta=0}\right)_{n,\veps} \beta = p_1 \beta \;.
\end{equation}

Note that, even though we have outlined the derivation in the simplest case (where the system is subthermal and close to equilibrium), the argument applies more generally. A broader discussion would rely on a detailed description of the out-of-equilibrium physics, which is not included in equation of state tables currently used for numerical simulations. 
Specifically, as described in \cref{app:compOSE}, starting from a three-parameter equation of state from the compOSE database, the derivatives we need can be worked out before carrying out a simulation. We are relying on this in the specific example discussed later.

\section{A multi-scale argument}\label{sec:Multiscale}

At this point we have formulated the relaxation problem for the reactive system. The equations from \cref{sec:hydrodynamics} are, in principle, all we need to evolve the system. 
However, it may not be numerically practical to solve the full nonlinear system, for example when the physical reactions are fast compared to numerically resolvable timescales. Given this, it is natural to consider approximations.

\subsection{The bulk-viscosity approximation}

To set up the discussion let us introduce the (proper) time derivative $d/dt = u^a\nabla_a$. We can then write the hydrodynamical equations in non-dimensional form (assuming for a moment that we work with the lepton fraction instead of $\beta$, an assumption that makes no practical difference here)%
\begin{subequations}
\begin{align}
    \frac{d\veps}{dt} &= - \frac{1}{\eps_{St}} (\veps + c_r^2 p )\theta \;, \\
    a_b &= - \frac{1}{\eps_{St}}\frac{1}{\eps^2_{Ma}} \frac{1}{\veps + c_r^2 p}\perp^{c}_b\nabla_c p \;,\\ 
    \frac{dn}{dt} &= -\frac{1}{\eps_{St}}n\theta \;,\\
    \frac{d Y_\e}{dt} &= -\frac{1}{\eps_A} (Y_\e - Y_\e^\mathrm{eq}) \;, 
\end{align}
\end{subequations}
where we have defined the dimensionless parameters 
\begin{equation}
    \eps_{St} = \frac{l_r}{u_r t_r} \;, \quad 
    \eps_{Ma} = \frac{u_r}{c_r} \;, \quad 
    \eps_{A} = \frac{1}{\A t_r} \;,
\end{equation}
and introduced a reference sound speed $c_r$ as well as $l_r, t_r, u_r$ as reference lengthscale, timescale and fluid velocity---so that $a_b$ has dimensions $u_r/t_r$. From this we see that $n,\veps, u^a$ evolve on similar timescales (in terms of the proper time associated with $u^a$), while the electron fraction evolution timescale is given by $\eps_A$. Assuming---as expected for large regions in neutron star merger simulations---that reactions occur on a fast timescale (so that $\eps_A \ll 1$), we may consider three different regimes: i) The expansion rate\footnote{As can be seen from e.g. \cref{eq:BetaEqGeneral}, the expansion rate is a ``source-term'' in the affinity equation.} $\theta$ varies only (or primarily) on slow timescales---which makes sense for numerical simulations where the spatial dynamics are resolvable but the reaction timescale is not; ii) The expansion $\theta$ varies on the fast timescale and hence this must be resolved in a simulation; iii) The flow is turbulent and therefore all scales are coupled.
In the last two cases, we cannot analytically simplify the problem much---expensive direct numerical simulations are required, although the large-eddy strategy \cite{mcdonough,lesbook,carrasco,LES1} (see also \cref{sec:LES}) may provide a useful alternative. As our main interest here is to consider the regime where progress can be made through approximations, we focus on the first case, where we can use standard multi-scale methods (see, e.g.~\cite{Weinan}) to ``integrate out'' the fast behaviour.

Bringing the multiscale argument from \cite{StuartPavliotis} to bear on the problem and assuming that we continue to work with $\beta$ (see \cref{app:MultiScale} for more detailed steps), we have to compute the late-time behaviour for the affinity by integrating the $\beta$ equation considering the other variables as parameters, and then taking the limit $t\to\infty$. The approximated equations for the remaining variables are then unchanged to lowest order, but we have to evaluate every function of $\beta$ using the late-time result. 
This is intuitively motivated by the underlying assumption that the affinity evolves on a faster timescale, so that the remaining degrees of freedom are approximately frozen on short timescales. 
The inclusion of the first order corrections then guarantees the approximated equations to be correct up to $\O(\eps_A^2)$ up to times $\O(1)$. The net result is that the pressure in the energy and number density, as well as in the Euler equation, will be approximated as
\begin{equation}
    p = p^\mathrm{eq} -\left(\frac{\partial p }{\partial \beta}\right)_{n,\veps}  \frac{n}{\A}\B\theta
\end{equation}
The ``interpretation'' of the second term as the Navier-Stokes bulk-viscosity is supported by the analysis in \cref{sec:thermodynamics} (cf. \cref{eq:Pi_t}). This result, while  in line with ``expectations'', is non-trivial. In fact, common arguments in favour of a representation of the net effect of under-resolved reactions via a bulk-viscous pressure are typically  perturbative in nature \cite{AlfordBulk10,AHSparticles20}.

In \cref{app:ElectronFraction} we provide the corresponding results obtained in terms of the electron fraction. The results agree---as they should---and the approximate equations we obtain by integrating out the electron fraction degree of freedom describe a Navier-Stokes bulk-viscous fluid, as well.
A relevant consequence of the electron fraction analysis from \cref{app:ElectronFraction} relates to the ``double counting'' issue raised in \cite{PeteThermal}. The discussion in \cite{PeteThermal} points out that the effects of under-resolved reactions are already accounted for in schemes aimed at modelling neutrinos, and hence adding a bulk-viscous pressure on top of that may lead to a double-counting.  Specifically, in \cref{appsub:doublecounting} we provide a proof of principle argument that supports this. We consider---even though this may be/sound somewhat ad-hoc---the case where we can split the reactions in two families,  fast ones (representative of, say, direct Urca processes) that cannot be captured by the numerics, and  slow ones (representative of, say, modificed Urca processes) that can be resolved. It would seem reasonable after the discussion we just had, to model the impact of unresolved reactions as a bulk-viscosity. Whilst this may be a valid strategy, one has to be careful because the multi-scale methods result of \cref{eq:DoubleCountingResult} suggests that the resolvable reaction rates should pick up a correction term as well.

\subsection{Making contact with simulations}

Having discussed the approximate equations we obtain from the multi-scales approach, it makes sense to ``step back'' and ask to what extent we expect this approximation to make sense for numerical simulations. To set the stage for the discussion, let us rewrite \cref{eq:BetaEquationSubthermal} as
\begin{equation}
    \frac{d\beta}{dt}= - \A \beta + \B \frac{d n}{dt} \; .
\end{equation}
The parabolic limit---which corresponds to neglecting the time derivative of the affinity $ d\beta / dt$---in the Fourier domain is
\begin{equation}
    \beta_{NS} (\omega)= \frac{\B}{\A} n\omega \;,
\end{equation}
while the extended irreversible thermodynamics (EIT) result takes the form
\begin{equation}
    \beta_{EIT} (\omega)= n\B\A \frac{\omega}{\omega^2 + \A^2} = \beta_{NS}(\omega) \frac{\A^2}{\A^2 + \omega^2} \; .
\end{equation}
\begin{figure}
\centering
\includegraphics[width=0.9\textwidth]{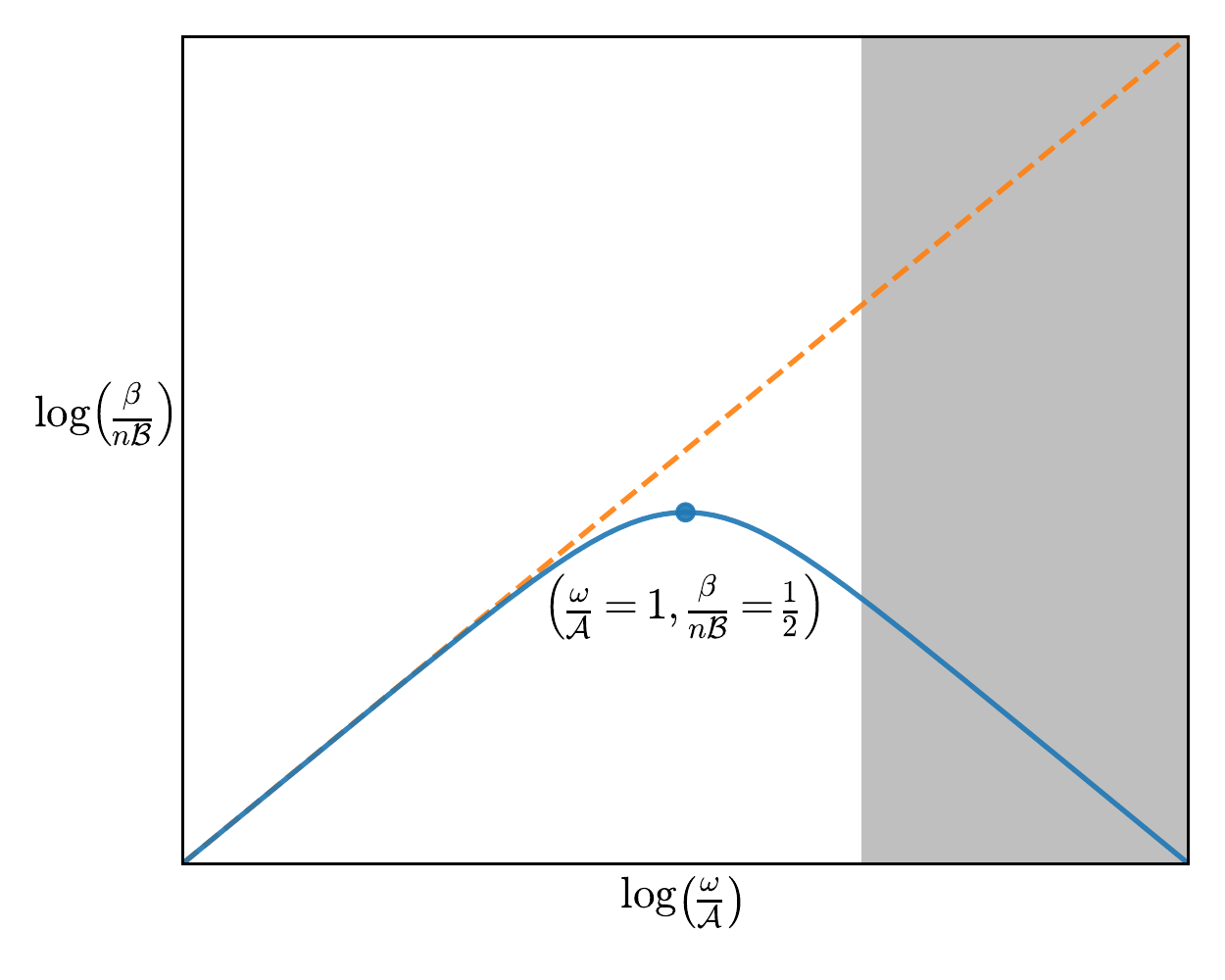}
\caption{Illustrating the behaviour of  $\beta(\omega)$ in two cases: the solution to the Cattaneo equation (i.e. the full extended irreversible thermodynamics (EIT) relaxation towards the Navier-Stokes limit, blue solid curve) and the  parabolic case (the Navier-Stokes limit, orange dashed line). The shaded region indicates  frequencies we assume we may not have ``access'' to numerically (this region  moves towards higher frequencies when the numerical resolution is increased).
}\label{fig:AffinityPlot}
\end{figure}
The two results are compared in \cref{fig:AffinityPlot}, from which we see that the EIT behaviour shows the ``expected'' resonance feature \cite{Sawyer89,AlfordBulk10,AlfordHarrisDamping18,AHSparticles20,AlfordHarutyunuanSedrakian19,AlfordHarutyunuanSedrakian21}. We note, however, that this is not the usual figure (see, for example, figure 2 in \cite{AlfordHarrisDamping18}), as we are plotting the affinity $\beta$ and not the bulk-viscosity coefficient. To link the results, we need to recall that $\Pi = p_1 \beta $ and then define $\zeta$ via  $ \Pi = - \zeta \theta$ . Then we can use \cref{eq:baryonCont} to write  $\theta = - {\dot n}/{n}$ to obtain
\begin{equation}
    \zeta_{NS} = p_1 \frac{\B}{\A} n  \;, \quad \text{and}\quad \zeta_{EIT} = \zeta_{NS} \frac{\A^2}{\omega^2 + \A^2} \; .
\end{equation}
The difference is subtle as both  $\beta_{EIT}$ and $\zeta_{EIT}$ present resonant features. However, $\zeta_{EIT}$ does so when the frequency $\om$ is kept fixed and $\A$ is varied, while $\beta_{EIT}$ exhibits the resonance even if we fix $\A$ and vary the frequency.

The difference may not seem particularly relevant, but the illustration in \cref{fig:AffinityPlot} allows us to draw useful conclusions. The figure shows that the parabolic limit (i.e. the limit we represent with the multi-scale argument) is a ``good'' approximation at low frequencies (see \cref{app:BoundTheError}), but becomes less accurate at higher frequencies\footnote{The illustration also provides an intuitive demonstration of why the high-frequency behaviour is non-causal and associated with a linear instability~\cite{HiscockInsta}.}. Keeping in mind that we consider the problem in the context of numerical simulations, we may also note that there will inevitably be frequencies that we do not have access to. This high-frequency cut-off is (schematically) represented by the shaded region in~\cref{fig:AffinityPlot}. Specifically, the resolution is limited by $\omega \sim \Delta t ^{-1}$ where $\Delta t $ is the numerical timestep. Now, because the resonance frequency is given by $\omega = \A$, we are left with two options: either the numerical timestep is large enough  ($\Delta t ^{-1} < \A$) that the peak is not resolved, or the simulation is precise enough that the region to the right of the resonant peak is (at least partially) resolved. In the first case, our simulation cannot resolve the (fast) relaxation towards the Navier-Stokes limit (and it is also likely that the expected instability \cite{HiscockInsta} associated with the Navier-Stokes behaviour will be suppressed). In the second case, the relaxation towards Navier-Stokes can be resolved by the numerics, so working with the Navier-Stokes approximation is just wrong.

\section{How relevant is bulk viscosity in mergers?}
\label{sec:relevance}

In the previous sections we have described how a reactive system near equilibrium can be approximated as a single fluid model with a bulk viscous pressure. One question to tackle is whether this approximation is of any relevance for, for instance, a neutron star merger simulation. In order for the argument to be of interest we need multiple criteria to be met. First, the timescales that we can resolve in the simulation must not include the key timescales for the reactions. If we can resolve the reactions then we should, as the bulk viscosity approximation (as well as the Cattaneo-type law \cref{eq:BetaEquationSubthermal}) only includes the linear effects, whilst the reactions account for the full nonlinear behaviour. Besides, the illustration in fig.~\ref{fig:AffinityPlot} shows that the approximation would simply break down in a resolved simulation. Second,  we would require that the correction to the total pressure from the bulk viscous approximation should not be negligible. If the bulk viscous term due to the reactions is tiny in comparison to the standard fluid pressure then it is pointless to include the reactions in the numerical simulation. Instead we should impose chemical equilibrium directly, and solve without reactions or a bulk viscosity approximation.

\begin{figure}
    \centering
    \includegraphics[width=\textwidth]{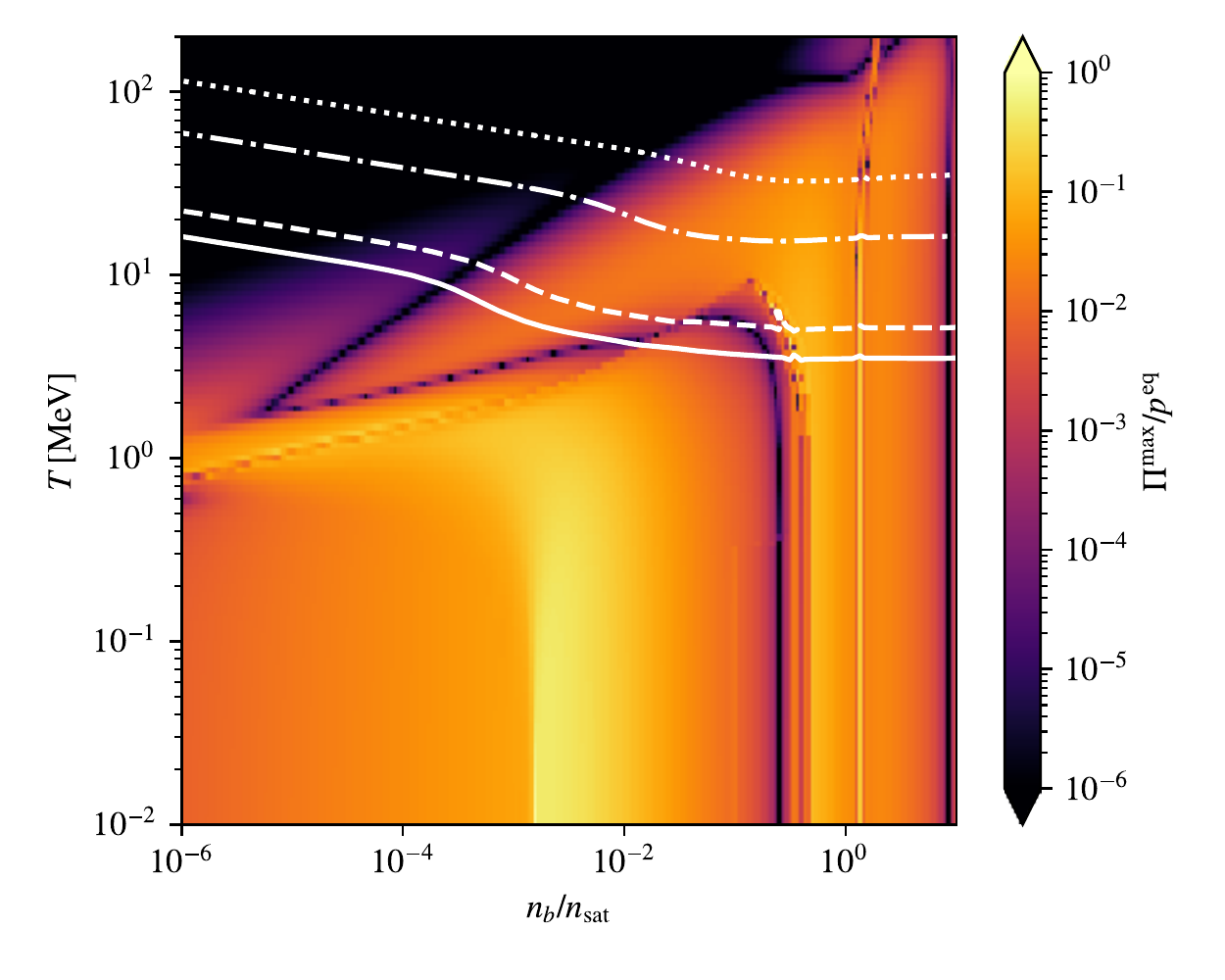}
    \caption{The maximum (for each point we assume that $\omega=\A$) potential relative contribution of the bulk viscous approximation $\Pi^\mathrm{max}/p^\mathrm{eq}$ at each point in phase space using the APR \cite{compOSE,APReos} equation of state.
    We see that the bulk viscous pressure contribution can be large for most temperatures when $n \gtrsim 10^{-3} n_{\text{sat}}$. However, the bulk viscous approximation should only be used where the reaction rate cannot be resolved by the numerical simulation, which is where the grid frequency is greater than $\A$. Also shown are contours at $\A=\{10^3,10^4,10^7,10^9\}~\mathrm{s^{-1}}$ (solid, dashed, dot-dash, dot). For current simulations, frequencies of $\sim 10^6~\mathrm{s^{-1}}$ are resolvable. This shows that the bulk viscous approximation should only be used for $T \gtrsim 10$MeV, and as the grid resolution improves becomes less necessary.}
    \label{fig:P_test_new}
\end{figure}

Figure~\ref{fig:P_test_new} combines both these criteria in a single plot (based on data for the APR equation of state \cite{APReos,compOSE} used in the simulations discussed in \cite{PeteThermal}). In order for the bulk viscous approximation to be useful the timescale from the simulation must be slower, or at a lower angular frequency, than that given by the peak frequency of the resonance. This peak is defined by $\A$. Therefore every point in the $(T, n)$ plane \emph{below} a contour of fixed $\A$ has reactions acting on time or frequency scales that can (and hence should) be resolved by the numerical simulation. The contours given show that as the numerical grid resolution is improved, more of the $(T, n)$ plane should be modelled by solving directly for the reactions. The bottom two contours in the figure bracket frequencies relevant for bulk gravitational wave generation ($1-10$kHz), and so must be captured by any numerical simulation. The top contour lines are indicative of the grid frequencies achievable by current simulations (an angular frequency of $10^7$s${}^{-1}$ roughly corresponds to a grid spacing of $200$m) and the state of the art in maybe ten years (an angular frequency of $10^9$s${}^{-1}$ roughly corresponds to a grid spacing of $2$m). However, there remain points at densities above $10^{-2} n_{\text{sat}}$ and temperatures above $10$MeV where reactions can only be modelled using the bulk viscous approximation, even with the best resolution currently available. Current simulations such as~\cite{PeteThermal} do see substantial amounts of matter in the post-merger remnant within this part of phase space, meaning that the bulk viscous approximation will remain necessary for the foreseeable future.

In addition, \cref{fig:P_test_new} shows the maximum relative magnitude of the bulk viscosity pressure, again in the $(T, n)$ plane, with contours of fixed $\A$ overlaid. In this calculation the bulk viscous approximation requires a range of additional approximations, including assuming the Fermi surface approximation is valid in order to compute the rates. These approximations are laid out in detail in \cref{app:compOSE}. However, for the purposes of our argument, the key is to note that we are interested in regions of phase space where the bulk viscosity approximation may be valid (above the contours), and where the bulk viscosity makes a significant contribution to the pressure. We see that this holds again in the region with densities above $10^{-2} n_{\text{sat}}$ and temperatures above $10$MeV.

As a final important point we note that there is no sharp transition between regions where the effects of the bulk viscous approximation are sizeable and where they are not, when considering contours of fixed $\A$ (which can be linked to the numerical resolution). Unless the numerical resolution can be increased by many orders of magnitude beyond the current state of the art, there will always be regions in spacetime where the bulk viscosity is significant but the approximation itself is debatable. Therefore we have to consider a model that is able to model reactions by directly evolving, for example, the species fraction in some parts of spacetime, makes the bulk viscous pressure approximation in other parts of the spacetime, and transitions between the two appropriately. This is difficult to do correctly.

\section{The impact of large-eddy filtering}\label{sec:LES}

We have demonstrated how the Navier-Stokes limit for bulk viscosity due to reactions can be obtained through a multi-scale argument, effectively integrating out the fast timescales of the problem. The argument is quite intuitive. However, the discussion is not yet complete. In the context of numerical simulations we also have to  consider other ``filtering'' aspects. In particular, we need to explore the link to (or conflict with) the large-eddy strategy. This problem is not straightforward. In a sense, the large-eddy approach is ``complementary'' to the approximate scheme as it aims to represent the regimes where multi-scale arguments do not  apply---namely turbulent flows and when all dynamics take place on fast timescales. 
\change{One may be tempted to view the multi-scale method and large-eddy space-filtering as ``orthogonal''. A space filter cuts off short length-scales, while  the invariant manifold method integrates out the fast dynamics (removing short timescales). 
However, the two issues are linked.} 

An actual numerical implementation introduces an implicit filtering associated with the discretized numerical grid. 
On a grid with fixed spacing $\Delta x$, the implicit filtering requires that any spatial feature on shorter lengthscales cannot be captured and must be modelled by some closure relation, as discussed in \cite{radice1,radice2,carrasco,viganoGR,duez}. Equally, the CFL bound (linked to causality) imposes that the timestep $\Delta t \propto \Delta x$, and so any physical feature happening on shorter timescales cannot be captured. Thus, increasing the accuracy by modifying the grid spacing automatically means the $\Delta t$ in \cref{fig:AffinityPlot} decreases and therefore the amount of the fast reactions that cannot be captured, shown by the grey area, decreases. There is a direct link between the amount of physics that must be modelled via a spatial filtering and via a time filtering or multi-scale argument.

In order to examine how the filtering associated with the large-eddy strategy impacts on the discussion of bulk viscosity, let us frame the argument using the ``fibration'' framework developed in \cite{LES1}. We refer to the original paper for a discussion of why---and under which assumptions---we may assume that the filtering operation (represented by angled brackets)  ``commutes'' with the covariant derivative, so that we have $\la\nabla_a A\ra = \nabla_a \la A\ra$.

Let us first consider  what happens when filtering is applied to the continuity equations for baryons and electrons, \cref{eq:baryonCont,eq:ElectronFractionEq}. As discussed in \cite{LES1}, the equation for the baryons remains unchanged if (and only if!) we choose to work with the density-weighted four-velocity:
\begin{equation}
    \nabla_a n^a = 0 \Longrightarrow \tu^a \nabla_a \tn + \tn\nabla_a \tu^a = 0 \;,
\end{equation}
where $\la n^a\ra = \tn \tu^a$ defines the filtered baryon number density current (indicated with a tilde). In the case of  the electrons, we need to introduce the ``electron fraction residual''
\begin{equation}
    \tau^a_{Y_\e} = \la Y_\e n^a \ra  - \la Y_\e\ra \la n^a\ra
\end{equation}
and work with a coarse-grained electron fraction,  defined by $\tn_\e = - \tu_a \la n_\e u^a\ra$ and leading to
\begin{equation}
    \tY_\e = \frac{\tn_\e}{\tn} = - \frac{\tu_a}{\tn} \la n_\e u^a\ra  \;.
\end{equation}
The filtered equation for the  electron fraction then becomes\footnote{Note that we could choose a different coarse-grained observer in such a way that the effective creation rate is re-absorbed in the macroscopic fluid four-velocity. However, this would come at a cost since the equation for the filtered baryon current would then have an additional diffusion term on the right-hand-side.}
\begin{equation}
    \tn \tu^a\nabla_a \tY_\e = \la\Gamma_\e\ra - \nabla_a \tnu_\e^a \;,
\end{equation}
where
\begin{align}
    \tY_\e = \la Y_\e \ra - \frac{1}{\tn} \tu_b\tau_{Y_\e}^b \;, \qquad \text{and } \tnu_\e ^a = \tperp^a_b \la n_\e u^b \ra =  \tperp^a_b\tau_{Y_\e}^b
\end{align}
The take home message is simple. The large-eddy filtering introduces an effective creation rate in addition to the faithful microphysical one. As the effect of the reactions can be modelled as a bulk-viscosity this may clearly have an impact on the analysis (cf. \cref{eq:BetaEqGeneral}). 

Now turn to the remaining equations of motion, e.g. \cref{eqs:Euler}, which follow from the conservation of the stress-energy-momentum tensor. As this retains exactly the same form as in the perfect fluid case, we can simply draw on the results from \cite{LES1}. The only difference is in the pressure and the Gibbs relation, as we are now considering a reactive system: 
\begin{equation}
    \la p \ra = \la - \veps + Ts + \mu_\n n - n_\e \beta \ra\;.
\end{equation}
The  $n_\e\beta$ term was not considered in \cite{LES1} as the fine-scale model considered there did not account for reactions. Nonetheless, we may simply adapt the same strategy: introduce an effective three-parameter equation of state at the resolved scale, and use it to define the macroscopic thermodynamic variables
\begin{subequations}
\begin{align}
    \frac{1}{\tT} &\doteq \left(\frac{\partial \ts}{\partial \tilde \veps}\right)_{\tn,\tY_\e}\;,\\
    - \frac{\tmu_\n}{\tT} &\doteq  \left(\frac{\partial \ts}{\partial  \tn}\right)_{\teps,\tY_\e} - \frac{\tY_\e}{\tn} \left(\frac{\partial \ts}{\partial  \tY_\e}\right)_{\tn,\,\teps}\;,\\
    \frac{\tbeta}{\tT} &\doteq \frac{1}{\tn}\left(\frac{\partial \ts}{\partial \tY_\e}\right)_{\tn,\teps} \;.
\end{align}
\end{subequations}
With these definitions in hand, we can  rewrite the filtered Gibbs relation as
\begin{equation}
    \la p \ra = -\teps + \tn \tmu_\n + \tT \ts - \tn\tY_\e\tbeta + M
\end{equation}
with the enhanced closure term
\begin{equation}\label{eq:FilteringOutEqPressure}
    M = \left[\left(\la n \mu_\n \ra  -  \tn\tmu_\n\right) + \left(\la Ts \ra - \tT\ts\right) - \left(\la \veps \ra - \teps \right) - \left(\la n_\e \beta\ra - \tn\tY_\e\tbeta \right)\right] \;.
\end{equation}
It makes sense to introduce the effective pressure $\tp = \la p \ra - M$ as this will satisfy a Gibbs relation of the pre-filtered form (but now in terms of the coarse-grained equation of state and the associated variables). Then, because the filtered energy and Euler equations will contain $\la p \ra$ (cf. the discussion in section V of \cite{LES1}), the $M$ term will enter the final equations as a correction to the pressure---effectively providing a bulk-viscous-like contribution. Let us stress that while we are only providing a minimal discussion of a large-eddy model, this is everything we need here. To make real progress we would need to introduce an explicit closure scheme and perform numerical experiments, both of which go beyond the scope of the present work. 

The essence of the argument is that the large-eddy filtering introduces an ``effective bulk-viscosity'' contribution to the coarse-grained equations. 
This happens in (at least) two ways: i) through the residual term $M$ stemming from filtering the Gibbs relation ii) by adding an effective creation rate, and the two effects are not (necessarily) linked as they depend on the introduced closure relations. 
In particular, the effective creation rate (which follows from the four-divergence of $\nu_\e^a$) affects the effective restoring term $\gamma$, and in turn $\A$---as is evident from inspection of \cref{eq:Adef}. 
As the resonance frequency in \cref{fig:AffinityPlot} is given by $\omega = \A$, and this essentially dictates whether or not the Navier-Stokes approximation is applicable, it makes sense to consider applying the filtering \underline{and} the multi-scale approach at the same time. 

We can intuitively see (and check explicitly) that the  coarse-grained equations will have a bulk-viscous contribution stemming from having integrated out the electron fraction degrees of freedom, and one from the filtering. The analytic expressions of these two terms depend on the order with which we take the steps: either we apply the multi-scale methods first and then filter, or the other way around. The results are unlikely to be the same. To see this, simply note that the multi-scale/invariant manifold method essentially boils down to an approximation of the equations around the equilibrium surface $Y_\e = Y_\e^\mathrm{eq}$. If we take the filtering step first, the notion of equilibrium changes---both because the (potentially different) equation of state is evaluated in terms of the coarse grained variables, and because of the effective rate. 
This highlights the importance of including all the relevant physics when constructing the closure terms in a large eddy model. This is problematic as the best closure terms require direct fine scale numerical simulations, and the analysis in this paper shows these expensive simulations need repeating each time additional physics is added. It is bound to be an expensive endeavour.

\section{Concluding remarks} \label{sec:conclusions}

Binary neutron star mergers offer a unique opportunity to explore several extremes of physics, but we need to significantly improve our numerical simulation technology if we want to realize the discovery potential of future gravitational-wave instruments (required to catch the high-frequency dynamics of the merger events). In particular, we need to make sure that efforts to infer the detailed physics are not stumped by systematic errors associated with the numerical implementation.
An important step towards realism involves dealing with nuclear reactions. Motivated by this, we have considered the issue of reactions from the perspective of numerical simulations (and the associated limited resolution), aiming to provide ``new insights on an old problem''. Specifically, we have discussed to what extent it makes sense to capture the net effect of reactions via a bulk viscosity prescription. Our analysis was based on standard multi-scale methods and provide an argument flagging the issue of possible ``double-counting'' (see \cite{PeteThermal} and \cref{appsub:doublecounting}). Our discussion also explicitly considered the issue of resolution limitations. In essence, we assessed the impact of the reaction timescales on the way we should  frame the modelling, and represent the net effect of reactions. 

Our key messages link the reaction timescales to the numerical (grid) timescales. When the reactions are slow the ground truth result is found by evolving the reactions directly. When the timescales are comparable, particularly when the physical timescales are (slightly) faster, this is not numerically practical. Instead the evolution system must be approximated. To leading order the reactions relax the system to equilibrium instantly. However, the error incurred is proportional to the ratio in the timescales. The first order (in the ratio of scales) approximation introduces correction terms that act as a bulk viscosity. \change{This demonstrates how  a multi-component reactive system can be approximated as a dissipative single fluid. 
This bulk viscous structure emerges regardless of whether we formulate the problem in terms of $\beta$ (which would be natural from a thermodynamics perspective) or $Y_\e$ (to connect more directly with simulations)}.

In a neutron star merger simulation the ratio of scales covers all ranges, with the scales being comparable (and hence a bulk viscosity approximation necessary) particularly in the core shortly after merger. We also show that the prescription for the bulk viscosity --- either algebraically in a ``Navier-Stokes'' like fashion, or by providing an equation of motion in a ``Cattaneo'' or ``Israel-Stewart'' like fashion --- is irrelevant, as they agree (to the order of the approximation) in the regime where the approximation is useful (at low frequencies where the ratio of scales is comparable). Finally, we show explicitly how the equations of motion are modified when some, but not all, of the scales are comparable. This flags up how the introduction of a bulk viscosity (either directly through approximating reactions, or indirectly through using large-eddy simulation techniques) can conflict with the definition of equilibrium. Consistently accounting for these issues to avoid ``double-counting'' is possible, \change{as outlined in appendix~\ref{appsub:doublecounting}}, but difficult. \change{A formulation that encompasses both fast and slow reaction rates will ultimately be required, but presents a challenge as it needs to consistently deal with the equilibrium and equation of state issues associated with the modified reaction rates seen in appendix~\ref{appsub:doublecounting}.}

Our main conclusions are intuitive but we believe this is the first time that they have been spelled out in detail. Contrasting theoretical work against the simulations in \cite{PeteThermal}, we find that different regions of the parameter space relevant for mergers would require different prescriptions. In particular, there are regions of the density-temperature phase space where reactions are slow enough that they can (and hence, should) be captured directly, and other regions where they are not---even with the best resolution available, now and in the foreseeable future---with no sharp boundaries between the two regimes. This indicates that, to properly account for reactions, we  need to develop numerical codes  capable of handling both regimes (reactions fast/slow compared to the resolved timescales), and the transition between them on the fly.

Finally, the issue of bulk viscosity is closely linked to the role of large-eddy filtering---which enters the discussion implicitly or explicitly. Noting this, we provided general filtering arguments that help set the stage for further work, \change{and highlighted the coupling between bulk viscosity and large-eddy modelling. A} ``definitive'' prescription for reactive systems will require explicit numerical experiments and the introduction of an appropriate closure model. 

\acknowledgments

We  are  grateful for support from STFC via grant numbers ST/R00045X/1 and ST/V000551/1.

\begin{appendix}

\section{Working with the CompOSE database}\label{app:compOSE}

Our analysis of the reactive problem is---ultimately---aimed at numerical implementations. Given this target, it makes sense to consider how the discussion impacts on the matter model that needs to provide the input physics. To be specific, we will spell out the connection with the compOSE database, which provides a useful collection of state-of-the-art equation of state models. As a result, the arguments draw heavily on the compOSE manual \cite{compOSE}, in particular  section 4.1.2 (``Thermodynamic consistency'') of version 2.0. The aim here is to explain how the various thermodynamical coefficients introduced in the main text can be worked out from an actual equation of state table. This is obviously a necessary step in the process. It also helps highlight to what extent existing tabulated data needs to be augmented in the future.

All equations of state relevant for our work in the compOSE database are provided as tables of $(T,n,Y_\q)$, where $Y_\q$ is the fraction of charged strongly interacting particles, which for a system without muons corresponds to the electron fraction $Y_\q = Y_\e$ (as local charge neutrality is assumed to hold). The central thermodynamical potential is the Helmholtz free energy density $f= \veps - Ts$, and some key quantities in the construction of the tables are 
\begin{equation}
     \left\{ \frac{p}{n},\, \frac{s}{n},\,\frac{\mu_\b}{m_\n}-1,\,\frac{\mu_\q}{m_\n},\,\frac{\mu_\l}{m_\n},\,\frac{f}{nm_\n}-1,\,\frac{\veps}{nm_\n}-1 \right\} \;,
\end{equation}
where $m_\n$ is the neutron mass---also provided in the tables and specific to each model---while $\mu_\b,\,\mu_\q,\,\mu_\l$ are the baryon, charge and lepton ``chemical potentials'' (respectively). The energy cost of adding a neutron, proton or electron to the system is then
\begin{equation}
    \mu_\n = \mu_\b \;,\quad \mu_\p = \mu_\b + \mu_\q \;,\quad \mu_\e = - \mu_\q + \mu_\l
\end{equation}
as follows straightaway from the respective baryon, charge\footnote{This is the total charge, not the charge of the strongly interacting particles corresponding to $n_\q = n Y_\q$.} and lepton number. The  baryon, charge and lepton chemical potentials are (in general) used to build the free energy, even though in the charge-neutral case with leptons this reduces to
\begin{equation}
    f = (\mu_\b  + \mu_\l Y_\e ) n - p \ .
\end{equation}
From this we see that there are only two independent chemical potentials at the thermodynamical level---consistent with a three-parameter equation of state---that can be written as derivatives of the Helmholtz free energy, namely $\mu_\b$ and $\mu_\l$.

Now, we need to connect with the quantities used in the main text. We have  (following from the Gibbs relation \cref{eq:OutEqGibbs})
\begin{equation}
    f = \veps - Ts \;, \quad df = -s dT + \mu_\n dn -\beta dn_\e
\end{equation}
where $\beta = \mu_\n - \mu_\p - \mu_\e$ in the cold equilibrium assumed throughout this paper. From this differential we see that, by definition 
\begin{subequations}
\begin{align}
    \mu_\n &= \left(\frac{\partial f}{\partial n}\right)_{T,n_\e} =  \left(\frac{\partial f}{\partial n}\right)_{T,Y_\e} - Y_\e \left(\frac{\partial f}{\partial n_\e}\right)_{T,n}\;, \\
    -\beta &= \left(\frac{\partial f}{\partial n_\e}\right)_{T,n} = \frac{1}{n} \left(\frac{\partial f}{\partial Y_\e}\right)_{T,n}\;.
\end{align}
\end{subequations}
Contrasting this with the results in section 4.1.2 of the compOSE manual, it is easy to see that\footnote{Note that $\mu_\q$ potential is not an independent thermodynamic quantity as the system cannot create protons alone because of the charge neutrality assumption.} $\mu_\n = \mu_\b$ and $\beta= - \mu_\l$. In practice, the affinity $\beta$ can be either read off from the tables directly, or computed as above. Because the equation of state table is  three-dimensional, the quantity extracted is inevitably a function of $\beta = \beta(n,T,Y_\q)$. This is consistent with the results of \cref{sec:hydrodynamics}, where we accounted for the fact that $\beta$ depends on either the temperature or the energy density.

Now, the coefficients $\A$ and $\B$ introduced in \cref{sec:hydrodynamics} can be obtained as combinations of derivatives of $\beta$ considered as a function of $(n,\veps,Y_\e)$. We need to link these expressions to derivatives that can be computed from the available tables (or, which may be more practical for the future, enhance the table with the required information). That is, we  have to change variables to arrive at
\begin{subequations}\label{eq:fromEtoT}
\begin{align}
    \left(\frac{\partial \beta}{\partial n}\right)_{\veps, Y_\e} &= \left(\frac{\partial \beta}{\partial n}\right)_{T, Y_\e} - \left(\frac{\partial \veps}{\partial n}\right)_{T, Y_\e} \left(\frac{\partial \veps}{\partial T}\right)_{n, Y_\e}^{-1} \left(\frac{\partial \beta}{\partial T}\right)_{n,Y_\e} \;, \\
    \left(\frac{\partial \beta}{\partial \veps}\right)_{n, Y_\e} &=  \left(\frac{\partial \veps}{\partial T}\right)_{n, Y_\e}^{-1} \left(\frac{\partial \beta}{\partial T}\right)_{n,Y_\e} \;, \\
    \left(\frac{\partial \beta}{\partial Y_\e}\right)_{\veps, n} &= \left(\frac{\partial \beta}{\partial Y_\e}\right)_{T, n} - \left(\frac{\partial \veps}{\partial Y_\e}\right)_{T, n} \left(\frac{\partial \veps}{\partial T}\right)_{n, Y_\e}^{-1} \left(\frac{\partial \beta}{\partial T}\right)_{n,Y_\e}  \;.
\end{align}
\end{subequations}
Recalling that these quantities should be evaluated at equilibrium, we see that we also need to construct the corresponding equilibrium table. Operationally, this can be done as follows.
We fix $n,T$ and vary $Y_\e$ until we find a value for which $\beta = 0$. The corresponding value of $Y_\e$ is then what we call $Y_\e^\mathrm{eq}$ and the equilibrium composition will automatically only be a function of $(n,T)$. Evaluating the original three-parameter model at $Y_\e = Y_\e^\mathrm{eq}$  gives the corresponding equilibrium energy density and pressure etc. Later in \cref{app:ElectronFraction} we derive an equivalent formula for $p_1$ that involves derivatives of $Y_\e^\mathrm{eq}$ with respect to the energy density $\veps$ and $n$. Using expressions analogous to \cref{eq:fromEtoT}, we can rewrite the result in terms of derivatives of $Y_\e$ with respect to $(n,T)$, that can be extracted from the tables.  

\begin{figure}
    \centering
    \includegraphics[width=\textwidth]{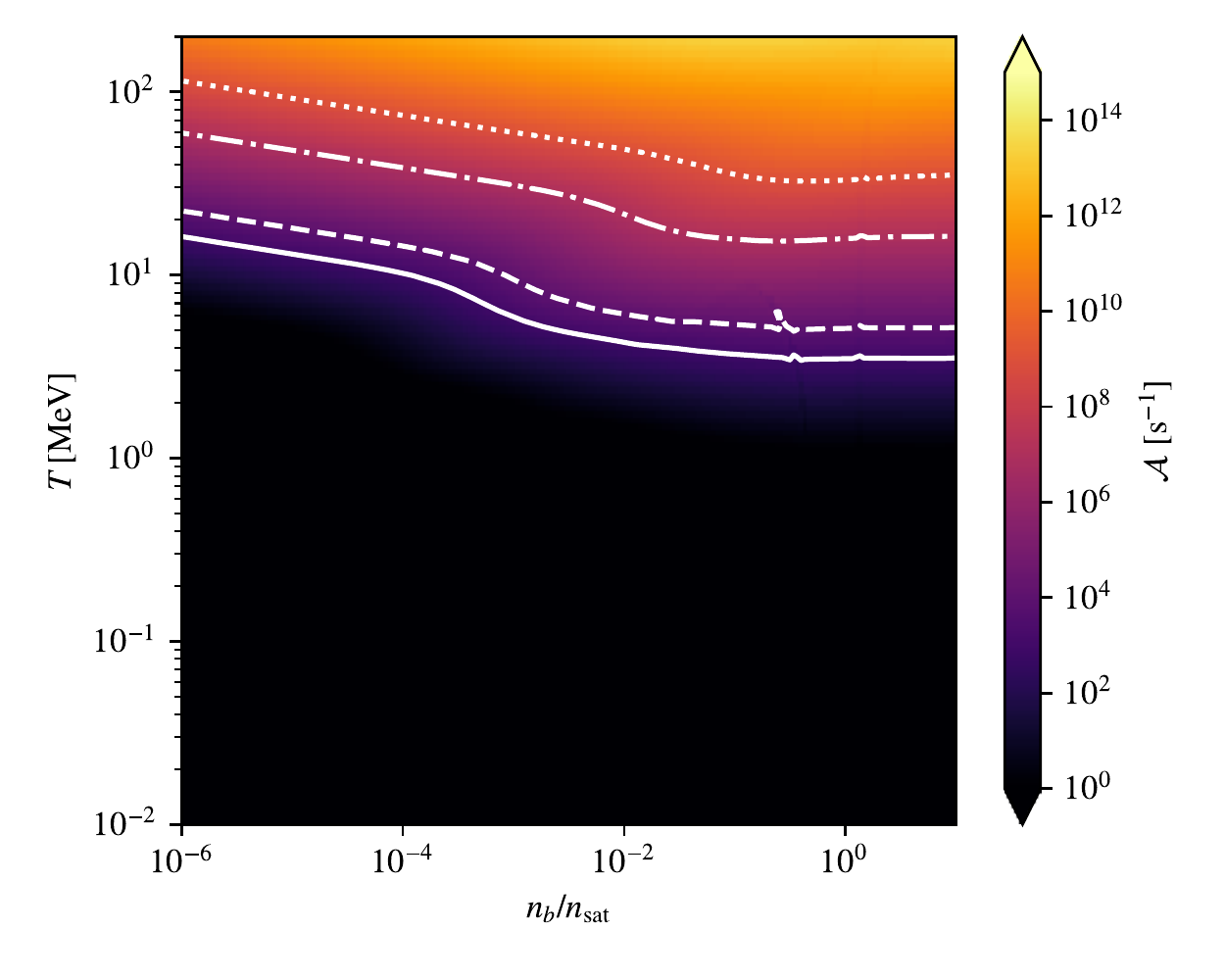}
    \caption{Plot of $\mathcal{A}$ for the APR equation of state used in \cite{PeteThermal}. The restoring term $\gamma$ is calculated assuming the Fermi surface approximation remains valid.  Contours are at $\mathcal{A}=\{10^3,10^4,10^7,10^9\}\mathrm{s^{-1}}$ (solid, dash, dot-dash, dot).}
    \label{fig:A_zoom}
\end{figure}

With the results in \cref{eq:fromEtoT} and evaluating the relevant quantities at equilibrium, we can work out (for a given equation of state) the value of $\B$, as required for \cref{fig:P_test_new}. In order to compute $\A$ though, we also need to evaluate the restoring term $\gamma$ (effectively, a measure of the reaction timescale). In \cref{fig:A_zoom} we show $\A$ as obtained from the modified Urca rates for the APR equation of state \cite{compOSE,APReos} used in~\cite{PeteThermal}. For this figure we have calculated $\gamma$ assuming the Fermi surface approximation, which allows us to use the analytic formulae from \cite{AlfordHarris18}.

\section{Neutrino-based estimate of the reaction timescale}

In appendix \ref{app:compOSE} we discuss the extraction of $\mathcal{A}$ from a three-dimensional tabulated equation of state (see figure \ref{fig:A_zoom}). This parameter is the inverse of the fluid equilibration timescale, and  we calculate it using the Fermi surface approximation for the reactions. The result should be valid for low temperatures ($T \lesssim 1~\mathrm{MeV}$), however the timescales relevant for our purposes occur in the range of (according to $\mathcal{A}$ as calculated above) $2~\mathrm{MeV}\lesssim T\lesssim 20~\mathrm{MeV}$ for the densities relevant to the neutron star core, begging the question of how accurate the approximation is.

Instead of calculating out-of-equilibrium rates without the Fermi surface approximation, we can take a different approach, and estimate the equilibration timescale using neutrino opacities, which can be readily calculated using software such as NuLib~\cite{NuLib}. The mean free path of the neutrinos before equilibration $\ell_\mathrm{eq}$ is then calculated from the absorption and scattering opacities, $\kappa_\mathrm{ab}$ and $\kappa_\mathrm{sc}$ respectively, through
\begin{align}
    \frac{1}{\ell_\mathrm{eq}} &= \kappa_\mathrm{eq} = \sqrt{\kappa_\mathrm{ab} \left(\kappa_\mathrm{ab} + \kappa_\mathrm{sc}\right)},
\end{align}
and we use the energy averaging method from~\cite{Endrizzi2020} to give $\tilde{\kappa}_\mathrm{eq}$, where the energy averaging is denoted by a tilde, and the corresponding $\tilde{\ell}_\mathrm{eq}=1/\tilde{\kappa}_\mathrm{eq}$. We then define an equilibration timescale $\tau_\mathrm{eq} = \tilde{\ell}_\mathrm{eq}/c$, by assuming the neutrinos are travelling at $c$. 

\begin{figure}
    \centering
    \includegraphics[width=\textwidth]{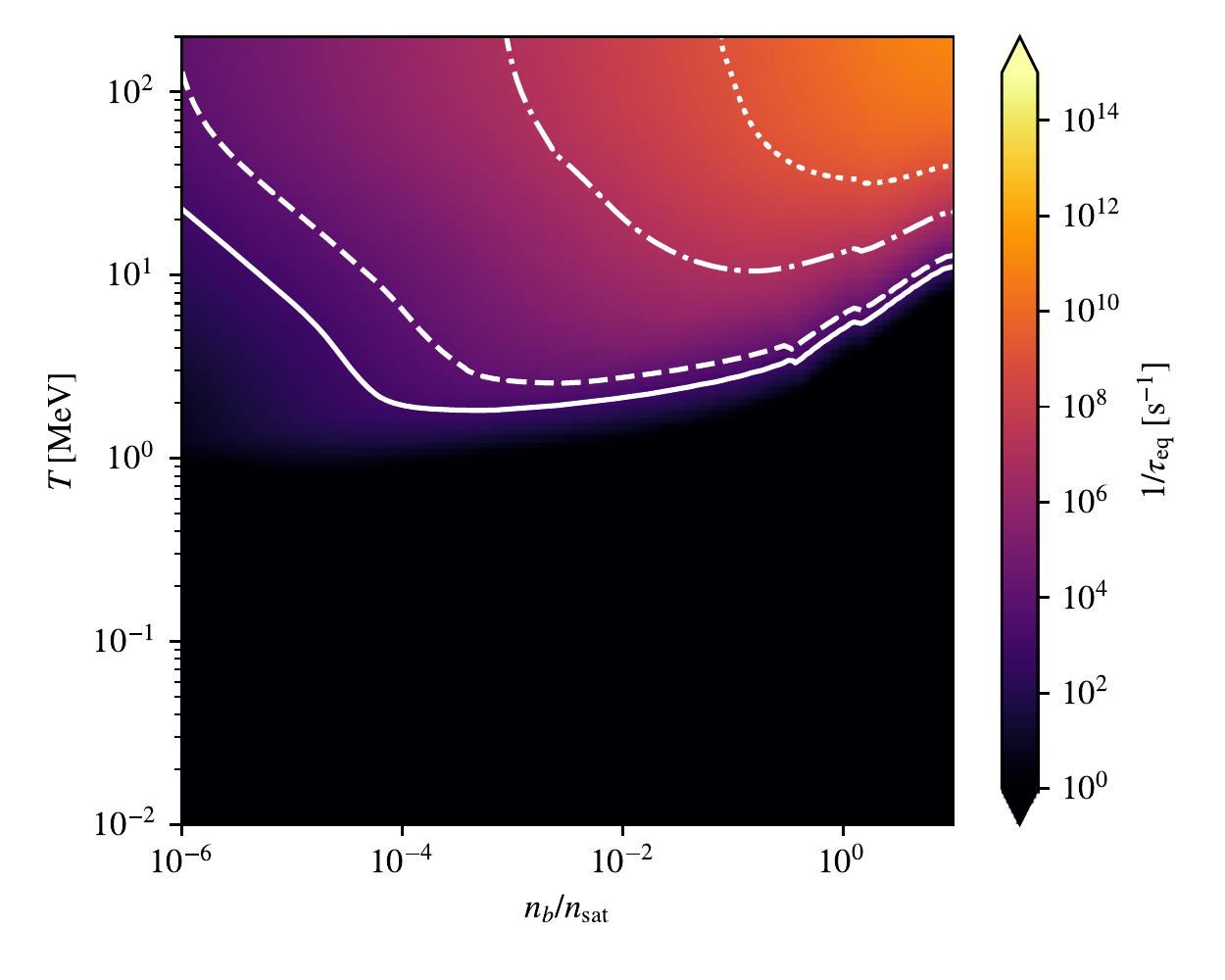}
    \caption{Plot of the inverse of the neutrino equilibriation timescale $1/\tau_\mathrm{eq}$. Contours are at $1/\tau_\mathrm{eq}=\{10^3,10^4,10^7,10^9\}\mathrm{s^{-1}}$ (solid, dash, dot-dash, dot).}
    \label{fig:neutrino_tau}
\end{figure}

In figure \ref{fig:neutrino_tau} we plot $1/\tau_\mathrm{eq} = \tilde{\kappa}_\mathrm{eq} c$ for the electron neutrinos, using the same fluid composition as in figure \ref{fig:A_zoom}. Comparing these results to figure \ref{fig:A_zoom} we see that there are some similarities. Both plots agree that the timescales for $T\lesssim 1~\mathrm{MeV}$ are too long to be relevant for our purposes, and they are also similar for most densities in the $1~\mathrm{MeV}\lesssim T\lesssim 10~\mathrm{MeV}$ region. However, at high temperatures there are  qualitative differences. We see the contours curving upwards much more noticeably here than for $\mathcal{A}$, and the timescales here are also generally slightly longer. We also note that at very high densities, where $n_b\gtrsim n_\mathrm{sat}$, there is an upwards trend to the contours, and they are also much closer together here than for the previous calculation. This suggests that an equivalent bulk viscous approximation will be relevant over less of the phase space when a timescale measure based on neutrino equilibriation is used compared to the Fermi surface approximation. The quantitative differences between the timescales shown in the two figures illustrates the uncertainties in the precise values due to not including all of the physics.

\begin{figure}
    \centering
    \includegraphics[width=\textwidth]{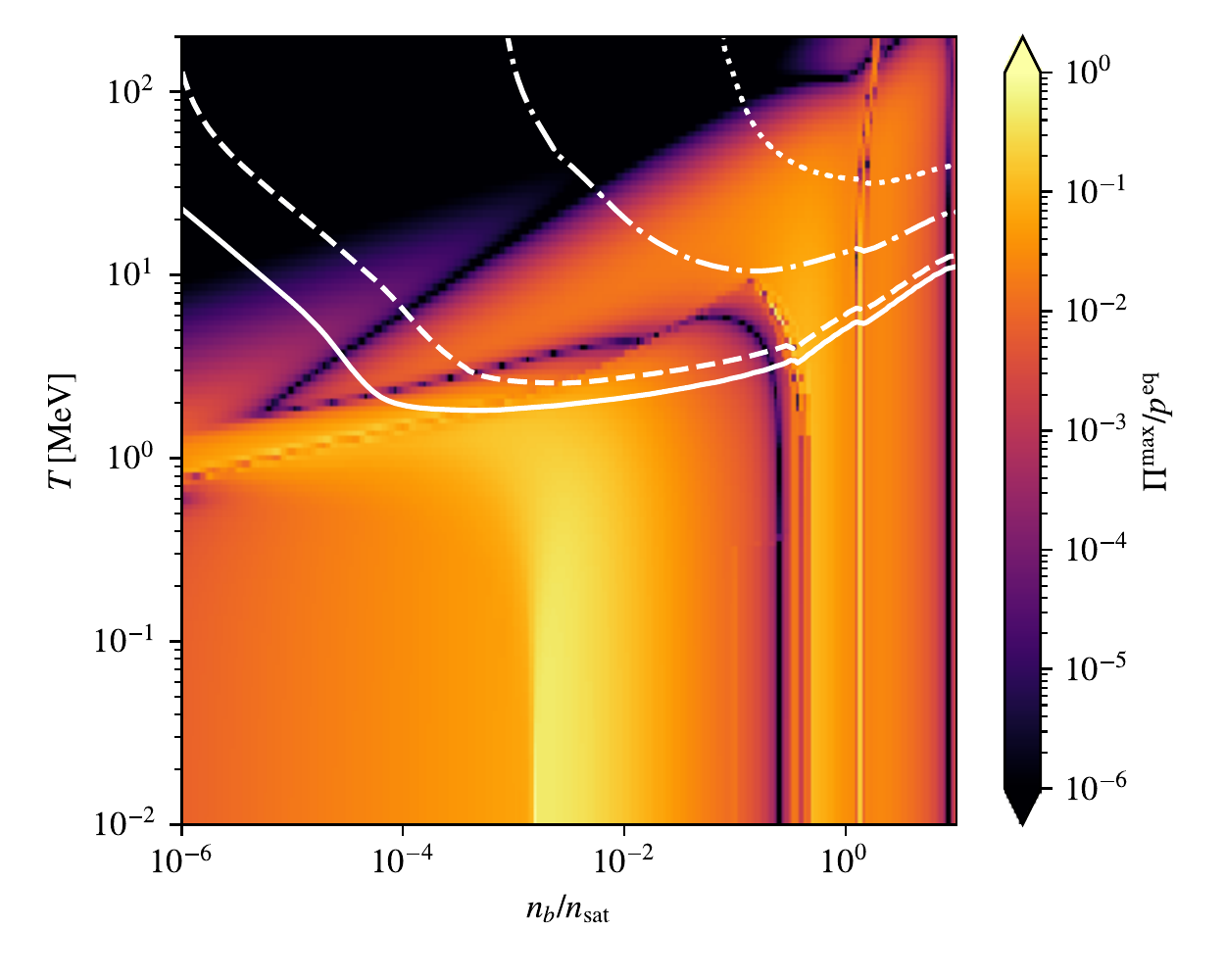}
    \caption{Copy of figure \ref{fig:P_test_new} with contours from figure \ref{fig:neutrino_tau}.}
    \label{fig:Pip_nu}
\end{figure}

Despite the qualitative differences between the equilibration timescales obtained through the two different methods, it is encouraging that there are some broad similarities between the two, particularly the regions of interest at relevant temperatures and densities. For completeness, we also include a version of figure \ref{fig:P_test_new} with contours from figure \ref{fig:neutrino_tau} as figure \ref{fig:Pip_nu}.

\change{\Crefrange{fig:P_test_new}{fig:Pip_nu} use the APR equation of state. We have in checked that analogous figures constructed using the SLy and SFHx equations of state are qualitatively similar, so that the key aspects of the discussion in this paper hold also for those equations of state. This suggests that the central results here are insensitive to the broad choice of equation of state.}

\section{Multi-scale arguments and the invariant manifold method}\label{app:MultiScale}

In order to support the arguments in the main text, we briefly summarize key results from~\cite{Weinan,StuartPavliotis}.

\subsection{Invariant manifold approach}

We assume we have a system of ordinary differential equations written as
\begin{subequations}
    \label{eq:multiscale}
    \begin{align}
        \dot{x} &= f(x, y), \label{eq:multiscale_slow} \\
        \dot{y} &= \epsilon^{-1} g(x, y). \label{eq:multiscale_fast}
    \end{align}
\end{subequations}
The variables $x$ are called \emph{slow}, and the variables $y$ are called fast, whilst $\epsilon \ll 1$ is a parameter.

In the \emph{invariant manifold} approach we assume that there exists an \emph{equilibrium (fast) state} $\varphi(x)$ such that
\begin{equation}
    \label{eq:multiscale_inv_manifold_equil}
    g(x, \varphi(x)) \equiv 0.
\end{equation}
We can then write the fast variables $y$ as an expansion in the small parameter $\epsilon$ about the equilibrium state as
\begin{equation}
    \label{eq:multiscale_inv_manifold_expansion}
    y = \varphi(x) + \epsilon y_1 + \mathcal{O}(\epsilon^2).
\end{equation}
Using the equations of motion~\cref{eq:multiscale} we find that the behaviour of the slow variables $x$ is approximated, to second order in $\epsilon$, by the solution to
\begin{subequations}
    \label{eq:multiscale_inv_manifold_solution}
    \begin{align}
    \dot{X} &= F_0(X) + \epsilon F_1(X) \\
    &= f(X, \varphi(X)) + \epsilon \nabla_y f(X, \varphi(X)) \left( \nabla_y g(X, \varphi(X)) \right)^{-1} \nabla_x \varphi(X) f(X, \varphi(X)).
    \end{align}
\end{subequations}

\subsection{Two timescale approach}

Strictly, the invariant manifold approach is only valid for ordinary differential equations. A more general approach that applies to partial differential equations is the two-scale approach. Using the two timescale approach as an example, this introduces the \emph{fast time} $\tau = t / \epsilon$ which is then treated as an independent variable. Applied to~\cref{eq:multiscale} this leads to
\begin{subequations}
    \begin{align}
        \partial_t x + \epsilon^{-1} \partial_\tau x &= f(x, y), \\
        \partial_t y + \epsilon^{-1} \partial_\tau y &= \epsilon^{-1} g(x, y).
    \end{align}
\end{subequations}
By gathering terms in powers of $\epsilon$, the fast behaviour can be integrated out by taking the integral average in $\tau$.

The result of the mathematical calculation is identical to the invariant manifold approach, when applied to ordinary differential equations. The calculation is general enough to include partial differential equations, and illustrates a different interpretation and potential problems. The interpretation is that the reduced system is valid for the \emph{integral average} of the slow variables: the fast scales have been integrated out. The potential problem is the requirement that the integral average of the fast behaviour is assumed to not contribute at leading order in $\epsilon$. This ``resonance'' behaviour cannot be captured by these approaches.

\subsection{Linear fast dynamics}
\label{sec:multiscale_linear}

A particularly relevant example is where the fast behaviour is linear, or can be linearised. In this case we write the full system~\cref{eq:multiscale} as
\begin{subequations}
    \label{eq:multiscale_linear}
    \begin{align}
        \dot{x} &= f(x, y), \label{eq:multiscale_linear_slow} \\
        \dot{y} &= \epsilon^{-1} (-A y + B), \label{eq:multiscale_linear_fast}
    \end{align}
\end{subequations}
with $A=A(x)$ and $B=B(x)$ are constants in the fast variables $y$. The equilibrium solution is therefore $\varphi(x) = B / A$, and the simplified system is
\begin{equation}
    \dot{X} = f(X, \varphi(X)) \left[ 1 + \epsilon A^{-1} \nabla_y f(X, \varphi(x)) \nabla_x \varphi(X) \right].
\end{equation}
The solution to the simplified system approximates the solution to the full system~\cref{eq:multiscale} to ${\mathcal O}(\epsilon^2)$ up to times ${\mathcal O}(1)$. The first order correction term needs to be applied consistently to all variables in the reduced system for this accuracy result to hold.

\subsection{Constructing the fast terms}
\label{sec:appendix_multiscale_construct_fast}

The construction in this section relies on the ratio of scales $\epsilon$ being explicit in the equations of motion. Usually a non-dimensionalisation of the system is needed to make the scales explicit. However, with complex nonlinear terms (such as tabulated net reaction rates which include many reaction channels) the precise form of the terms may not be obvious.

Here we need only the leading order terms and so can proceed as follows. We start from a system of equations which we expect to have fast behaviour
\begin{equation}
    \dot{z} = h(x, z),
\end{equation}
where $x$ are any variables we expect to be slow. We assume that we know how $h$ scales asymptotically with the ratio of scales. That assumption means we can explicitly compute
\begin{equation}
    h^{\text{fast}} = \lim_{\epsilon \to 0} \left( \epsilon h \right).
\end{equation}
This defines the source term for the fast behaviour as the piece that diverges linearly with the ratio of scales in the limit of infinitely fast speeds. We then split the source into fast and slow pieces using
\begin{equation}
    h^{\text{slow}} = h - \epsilon^{-1} h^{\text{fast}},
\end{equation}
and perform the equivalent split on the variables $z$ as
\begin{subequations}
    \begin{align}
        \dot{z}^{\text{fast}} &= \epsilon^{-1} h^{\text{fast}}, \\
        \dot{z}^{\text{slow}} &= h^{\text{slow}}.
    \end{align}
\end{subequations}
We can then identify the fast variables $y$ with ${z}^{\text{fast}}$ and augment the slow variables $x$ with ${z}^{\text{slow}}$.

\subsection{The reactive system}\label{app_subsec_reactive_system}

The equations of motion using the affinity can be written
\begin{equation}
    \frac{d}{dt} \begin{pmatrix} n \\ \veps \\ \beta \end{pmatrix} = \begin{pmatrix} -n \theta \\ -(p+\veps) \theta \\ - \A \beta - n \B \theta \end{pmatrix}.
\end{equation}
The dimensional analysis of~\cref{sec:Multiscale} indicates that the reaction timescale $\epsilon_A$ enters through $\A$ and through no other parameter---and it can be easily seen that $\B$ relates to how quickly the equilibrium electron fraction adjusts to a change in number and energy density.  Assuming only $\A$ to be fast, and using~\cref{sec:appendix_multiscale_construct_fast} we split the affinity into fast and slow pieces, writing
\begin{subequations}
    \begin{align}
        \frac{d}{dt} \begin{pmatrix} n \\ \veps \\ \beta^{\text{slow}} \end{pmatrix} &= \begin{pmatrix} -n \theta \\ -(p+\veps) \theta \\ - n \B \theta \end{pmatrix}, \\
        \frac{d}{dt} \begin{pmatrix} \beta^{\text{fast}} \end{pmatrix} &= \begin{pmatrix} - \epsilon_A^{-1} \A \beta \end{pmatrix},
    \end{align}
\end{subequations}
where $\beta^{\text{fast}} + \beta^{\text{slow}} = \beta$. By construction $\A$ is independent of $\beta$, so the fast dynamics are now linear and the results of~\cref{sec:multiscale_linear} apply. We see that the invariant manifold is given by $\varphi(X) = -\beta^{\text{slow}}$, which is equivalent to saying $\beta = 0$ on the invariant manifold. That is, the invariant manifold corresponds to beta equilibrium, as expected. 

The reduced system is
\begin{equation}
    \frac{d}{dt} \begin{pmatrix} n \\ \veps \\ \beta^{\text{slow}} \end{pmatrix} = \begin{pmatrix} -n \theta \\ -(p+\veps) \theta \\ - n \B \theta \end{pmatrix} - \epsilon_A \frac{n \B \theta}{\A} \begin{pmatrix} 0 \\ \frac{\partial p}{\partial \beta} \theta \\ 0 \end{pmatrix}\ ,
\end{equation}
where all terms (pressure and its derivatives, and $\B$) have to be evaluated at beta equilibrium. The final equation decouples as all quantities in the first two equations depend on the total $\beta$ evaluated at equilibrium, which is $\beta=0$.
We see from the energy equation that the pressure correction appears as
\begin{equation}
    \Pi_d = -\epsilon_A \left( \frac{\partial p}{\partial \beta} \right)_{n, \epsilon} \frac{n \B \theta}{\A} = - \left( \frac{\partial p}{\partial \beta} \right)_{n, \epsilon} \frac{n \B \theta}{\A},
\end{equation}
where in the second equality we have re-absorbed the scaling parameter into the reaction rate $\eps_A^{-1}\A \to \A$.

If we instead make the assumption that $\B$ is also a fast parameter, then it is easy to see that the affinity would be an entirely fast variable, that is $\beta^{\rm{slow}} = 0$ automatically. Repeating the analysis, the invariant manifold would now be $\varphi(X) = - n \B \theta / \A$, and as a consequence the pressure in the energy (as well as Euler) equation 
\begin{equation}\label{eq:expPbeta}
    p = p \left(n,\veps, - \frac{n \B\theta}{\A}\right) = p^\mathrm{eq} + \Pi_d \;.
\end{equation}
This shows that, in the case when  $\B$ is fast the bulk-viscous correction enters already at lowest order. The  corrections to this are second order in $\beta$, beyond the  regime of validity of the theory (linear in $\beta$), and hence cannot be trusted. In essence, whatever assumption we make for  $\B$ (being fast or slow), the bulk viscosity result is the same.

\section{Integrating out the electron fraction}\label{app:ElectronFraction}

\subsection{Revisiting the thermodynamics}\label{app:subsec:Yethermo}

As  explained in \cref{sec:hydrodynamics}, where we developed the bulk-viscosity argument in terms of the affinity $\beta$, we may equivalently work with the electron fraction $Y_\e$. Given this, we want to revisit the  path that we followed in the main text, and run through the two-timescale argument again, now  working with the electron fraction. As we will see, this also results in a correction term to the pressure, which will be naturally expressed in terms of derivatives involving the equilibrium electron fraction. Given this, it is natural to set the stage by revisiting the relevant thermodynamics results.

As the affinity $\beta$ is (roughly speaking) the thermodynamically conjugate variable to the electron fraction, it makes intuitive sense to start by considering the following thermodynamical potential 
\begin{equation}
    g = s - n_\e \frac{\beta}{T}
\end{equation}
such that 
\begin{equation}
    dg = \frac{1}{T}d\veps - \frac{\mu_\n}{T}dn - n_\e d\left(\frac{\beta}{T}\right)
\end{equation}
Exactly as we did for the entropy in \cref{sec:thermodynamics}, we can then expand $g$ around equilibrium. There will now be a first order term in $\beta$, which provides the ``formal definition'' of the equilibrium electron number density, $n_\e^\mathrm{eq}$. We get
\begin{equation}
    g(n,\veps,{\beta}/{T}) = s^\mathrm{eq} (n,\veps) - n_\e^\mathrm{eq}\frac{\beta}{T} + \frac{1}{2} g_2 \left(\frac{\beta}{T}\right)^2 \;,
\end{equation}
where
\begin{equation}
    n_\e^\mathrm{eq} = \frac{\partial g}{\partial ({\beta}/{T})}\Big|_{\beta = 0} \;,\quad\text { and }\quad g_2 = \frac{\partial ^2 g}{\partial \left( {\beta}/{T}\right)^2} \Big|_{\beta = 0}
\end{equation}

We can then work out the expansion (to first order in $\beta$) for $n_\e$ as 
\begin{equation}
    n_\e (n,\veps,\beta) = n_\e^\mathrm{eq} (n,\veps) - g_2 \frac{\beta}{T}    
\end{equation}
and use it in the definition of $g$ to arrive at 
\begin{equation}
    g = s^\mathrm{eq} + \frac{1}{2}s_2\beta^2 - n_\e^\mathrm{eq}\frac{\beta}{T}  + g_2 \left(\frac{\beta}{T} \right)^2 \;.
\end{equation}
Comparing this with the expansion above we identify
\begin{equation}
    g_2(n,\veps) = - T_\mathrm{eq}^2 s_2 = -  T_\mathrm{eq}\left(\frac{\partial \beta}{\partial n_\e}\right)_{n,\veps}^{-1} 
\end{equation}
Using this result and working out the expansion for the thermodynamical variables from $g$, we obtain (after linearizing in $\beta$ and  introducing $Y_\e^\mathrm{eq} = n_\e^\mathrm{eq} / n$) 
\begin{subequations}\label{eq:ExpansionsWithxp}
\begin{align}
    T &= T^\mathrm{eq} + n \left(\frac{\partial Y_\e^\mathrm{eq} }{\partial \veps}\right)_n T^\mathrm{eq} \beta \;, \\
    \mu_\n &= \mu^\mathrm{eq} + \left[\mu^\mathrm{eq}n \left(\frac{\partial Y_\e^\mathrm{eq} }{\partial \veps}\right)_n + n\left(\frac{\partial Y_\e^\mathrm{eq} }{\partial n}\right)_{\veps} - Y_\e^\mathrm{eq} \right]\beta \;,\\
    Y_\e &= Y_\e^\mathrm{eq} - \left(\frac{\partial \beta }{\partial Y_\e}\right)^{-1}_{n,\veps}\beta \;.
\end{align}
\end{subequations}
Combining the result with the Gibbs relation we can work out the pressure (and the ``thermodynamical'' bulk viscosity)
\begin{equation}
    p(n,\veps,\beta) = p ^\mathrm{eq}(n,\veps) + p_1 \beta = p^\mathrm{eq} + \Pi_t 
\end{equation}
with 
\begin{equation}\label{eq:p1Withxp}
    p_1 =  n \left[(p^\mathrm{eq} + \veps) \left(\frac{\partial Y_\e^\mathrm{eq} }{\partial \veps}\right)_n + n \left(\frac{\partial Y_\e^\mathrm{eq} }{\partial n}\right)_\veps\right] \;.
\end{equation}
The take home message is that we have  two equivalent expressions for the bulk-viscosity purely from thermodynamical arguments: this is always written as $\Pi_t = p_1 \beta$, where $p_1$ can be either written as in \cref{eq:expansionsWithBeta} or  \eqref{eq:p1Withxp}. These results are thermodynamically correct as long as the linearization in $\beta$ is valid (i.e., the system is close to chemical equilibrium), independently of the modelling of the relaxation towards equilibrium. 

\subsection{Invariant manifold method with the electron fraction}

Let us now run through the invariant manifold argument working with the electron fraction. The first step is to rewrite the equation of motion for the electron fraction as
\begin{equation}
    u^a\nabla_a Y_\e = \frac{\Gamma_\e}{n}= - \A (Y_\e - Y_\e^\mathrm{eq})
\end{equation}
where we have first taken the sub-thermal limit $\Gamma_\e = - \gamma \beta$ and then expanded around equilibrium\footnote{Let us note that, as is clear from the last of \cref{eq:ExpansionsWithxp}, $Y_\e = Y_\e^\mathrm{eq}$ if and only if $\beta = 0$, and that in the sub-thermal limit this means $\Gamma_\e (Y_\e = Y_\e^\mathrm{eq})= 0$. }. We then assume, as in \cref{sec:Multiscale}, that the electron fraction evolves on faster timescales than the other variables. 
This means we are effectively considering the system of equations
\begin{subequations}
\begin{align}
    \frac{d}{dt} \begin{pmatrix} n \\ \veps \end{pmatrix} &= - \begin{pmatrix} n\theta  \\ (p+\veps)\theta \end{pmatrix} \;, \\
    \frac{d}{dt} \begin{pmatrix} Y_\e \end{pmatrix} & = - \begin{pmatrix} \eps^{-1} \A(Y_\e - Y_\e^\mathrm{eq}) \end{pmatrix} \;.
\end{align}
\end{subequations}
where we have made explicit that the fast behaviour arise from $\A$ as explained in \cref{sec:appendix_multiscale_construct_fast}.
Again, the linear fast case of~\cref{sec:multiscale_linear} is relevant, giving that the invariant manifold is the equilibrium surface.
This immediately tells us that, to lowest order, the approximated equations describe a reactive fluid for which chemical equilibrium is restored immediately on the dynamical timescale. Including the first order corrections to the approximated equations---simply drawing on the results from \cite{StuartPavliotis} and \cref{sec:multiscale_linear}---we obtain
\begin{subequations}
\begin{align}
    \frac{d}{dt} \begin{pmatrix} n \\ \veps \end{pmatrix} &= - \begin{pmatrix} n\theta \\
    (p^\mathrm{eq} + \veps + \Pi_d) \theta \end{pmatrix}
\end{align}
\end{subequations}
where 
\begin{equation}\label{eq:BulkPressure_xeq}
    \Pi_d = \frac{n}{\gamma} \left(\frac{\partial \beta}{\partial Y_\e}\right)_{n,\veps} ^{-1} \left(\frac{\partial p}{\partial Y_\e}\right)_{n,\veps} \left[(p^\mathrm{eq} + \veps) \left(\frac{\partial Y_\e^\mathrm{eq} }{\partial \veps}\right)_n + n \left(\frac{\partial Y_\e^\mathrm{eq} }{\partial n}\right)_\veps\right]\theta
\end{equation}
Now, using the fact that 
\begin{equation}
    \left(\frac{\partial p}{\partial \beta}\right)_{n,\veps} = \left(\frac{\partial p}{\partial Y_\e}\right)_{n,\veps}\left(\frac{\partial \beta}{\partial Y_\e}\right)_{n,\veps}^{-1} 
\end{equation}
along with the two alternative formulae for $p_1$ \cref{eq:p1Withxp,eq:p1WithBeta} we observe the (pleasing) consistency with the result in \cref{app_subsec_reactive_system}. Exactly as in \cref{sec:Multiscale}, by integrating out the fast variable we pick up an additional contribution to the pressure that corresponds to a bulk-viscous response. 

\subsection{Partially resolved reactions and double counting}\label{appsub:doublecounting}

Let us now consider the situation where some part of the reactions are slow enough that we can capture them, and the rest is not. As an example, we may think the separation between the modified and direct Urca processes. Even though this might be somewhat artificial, the key point is that  we assume a clear scale-separation between  two types of reactions. We do so as we are only interested in a proof of principle argument here. We want to show the problems in constructing a consistent bulk viscous approximation in this case.

Let us start from the equation for the electron fraction, written in terms of the reaction rate. We define the fast/slow part of the total creation rate as discussed in \cref{sec:appendix_multiscale_construct_fast}
\begin{equation}
    \Gamma_f = \lim_{\eps\to 0} ( \eps\Gamma_\e) \;, \qquad  \Gamma_s = \Gamma_\e - \frac{1}{\eps}\Gamma_f \;,
\end{equation}
and hence split the electron fraction into its fast and slow contributions $Y_\e = Y_s + Y_f$ according to
\begin{equation}
    \frac{d}{dt} Y_s = \frac{\Gamma_s}{n} \;,\quad \text{and } \frac{d}{dt} Y_f = \frac{1}{\eps} \frac{\Gamma_f}{n} \; .
\end{equation}
We can think of $\Gamma_f$ as a function of $(n,\veps, Y_s+Y_f)$ and define a fast equilibrium fraction $Y_f^\mathrm{eq}$ such that $\Gamma_f(n,\veps,Y_s +Y_f^\mathrm{eq}) = 0$. It is then possible to expand the equation for the fast variable as
\begin{equation}
    \frac{d}{dt} Y_f = \frac{1}{\eps} \frac{\partial \Gamma_f}{\partial Y_\e}\Big|_{Y_\e =Y_s + Y_f^\mathrm{eq}} (Y_f - Y_f ^\mathrm{eq}) \; .
\end{equation}
Note that, because the two fractions must add up to the total electron fraction, we have $Y_f^\mathrm{eq}  = Y_\e -  Y_s$, namely the equilibrium fast fraction $Y_f^\mathrm{eq} = Y_f^\mathrm{eq} (n,\veps,Y_s)$. As for the slow reaction rate, we do not need to expand it around equilibrium because this is assumed to be resolved in the simulation. 
Applying the results of \cref{app:MultiScale} (including the first order corrections) we obtain
\begin{equation}
    \label{eq:DoubleCountingResult}
    \frac{d}{dt} \begin{pmatrix} n \\ \veps \\ Y_s \end{pmatrix} =  \begin{pmatrix} - n\theta  \\
    - (p + \veps + \Pi_d) \theta \\
    \Gamma_s -  \frac{1}{n} \left(\frac{\partial \Gamma_s}{\partial Y_\e}\right)_{n,\veps}\left(\frac{\partial p}{\partial Y_\e}\right)_{n,\veps}^{-1}\Pi_d  \end{pmatrix},
\end{equation}
with 
\begin{equation}
    \Pi_d =\left\{ \left[(p + \veps) \left(\frac{\partial Y_f^\mathrm{eq} }{\partial \veps}\right)_{n,Y_s} + n \left(\frac{\partial Y_f^\mathrm{eq} }{\partial n}\right)_{\veps,Y_s}\right]\theta  - \left(\frac{\partial Y_f^\mathrm{eq}}{\partial Y_s}\right)_{n,\veps} \frac{\Gamma_s }{n} \right\} \left(\frac{\partial \Gamma_f }{\partial Y_\e}\right)_{n,\veps}^{-1} \left(\frac{\partial p }{\partial Y_e}\right)_{n,\veps} \;,
\end{equation}
and everything is evaluated at $Y_f = Y_f^{\rm{eq}}$.
This argument then shows that, if there are both fast and slow reactions in the system, and we are trying to capture the effect of the fast/unresolved ones via a bulk-viscosity like contribution, we  need to tread carefully, as the introduction of the bulk viscosity also impacts on the resolved reaction rates, and the rates impact on the definition of equilibrium. This argument provides a quantitative demonstration of the ``double counting'' issue raised in~\cite{PeteThermal}.

\section{What bulk viscous pressure approximation is suitable?}\label{app:BoundTheError}

In the regime where the reactions need to be approximated as a bulk viscous pressure there are many possible ways in which such an approximation can be constructed. The standard first order form could be used, where the bulk pressure depends on a (data dependent) coefficient multiplied by the expansion associated with the fluid motion \cite{Most2021BV}. Alternatively, in frequency space the bulk pressure can be written as a term depending on the thermodynamics multiplied by a function of frequency. In this case the ``true'' (Cattaneo) result differs from the first order (``Navier Stokes'') result only in the form of the frequency term.

Intuitively (see the discussion around~\cref{fig:AffinityPlot}) we have argued that these two different bulk viscous approximations should be close to each other, as long as we are considering frequencies below the resonant peak. Here we make that argument more quantitative.

\subsection{Limiting frequency space}

We are interested in the low frequency part of the pressure that can be captured in a numerical simulation. We will therefore assume that we are only interested in frequencies $\wh < 1$ (``to the left of the peak''), and that there is a hard cut-off at $\whd \sim 2 \pi / (\A \, \Delta t)$ (the numerical scheme is spectral like and captures all frequencies available on the grid). Therefore
\begin{equation}
    \label{eq:bulk_pi_freq}
    \Pi = \int_{-\infty}^{\infty} \dd{\wh} H \left( 1 - \left| \frac{\wh}{\whd} \right| \right) p_1 \beta = 2 \int_{0}^{\whd} \dd{\wh} p_1 \beta.
\end{equation}

\subsection{Bounding the difference}

We want the relative difference between the two bulk viscous pressure approximations, which is
\begin{equation}
    \E = \frac{\left| \Pi_{\text{Cattaneo}} - \Pi_{\text{NS}} \right|}{\left| \Pi_{\text{NS}} \right|}.
\end{equation}
To compute this we \emph{assume} that we can bound the correction terms as powers of the frequency, as
\begin{equation}
    \label{eq:bound_coeffs}
    C_{-} \wh^a < n \B p_1 < C_{+} \wh^b,
\end{equation}
where $0 > a > b > -2$ is needed for the results to converge. This seems reasonable for Kolmogorov turbulence, but the range of these coefficients ($C_{\pm}, a, b$) has an impact on the result.

From this assumption we have
\begin{equation}
    2 \int_{0}^{\whd} \dd{\wh} C_{-} \wh^{1+a} < \left| \Pi_{\text{NS}} \right| < 2 \int_{0}^{\whd} \dd{\wh} C_{+} \wh^{1+b},
\end{equation}
giving
\begin{equation}
    \label{eq:bulk_NS_bound_both}
    \frac{2 C_{-}}{2 + a} \whd^{2+a} < \left| \Pi_{\text{NS}} \right| < \frac{2 C_{+}}{2 + b} \whd^{2 + b}.
\end{equation}

To bound the pressure difference, first write
\begin{subequations}
    \begin{align}
        \left| \Pi_{\text{Cattaneo}} - \Pi_{\text{NS}} \right| &= \left| n \B p_1 \wh \left( \frac{\wh^2}{1 + \wh^2} \right) \right| \\
        &= \left| n \B p_1 \wh^3 \right| + {\cal O}(\wh^5).
    \end{align}
\end{subequations}
From this we get
\begin{equation}
    \label{eq:bulk_NS_bound_diff}
    \frac{2 C_{-}}{4 + a} \whd^{4+a} < \left| \Pi_{\text{Cattaneo}} - \Pi_{\text{NS}} \right| < \frac{2 C_{+}}{4 + b} \whd^{4 + b}.
\end{equation}

Using the appropriate bound for numerator and denominator in $\E$, it follows that
\begin{equation}
    \label{eq:bulk_error_bound}
    \E < \frac{C_{+}}{C_{-}} \frac{2 + a}{4 + b} \whd^{2 + b - a}.
\end{equation}

\subsection{Interpretation}

As $a > b$ and $a-b < 2$ we see the difference between the two approximations is, in the frequency range ($\whd < 1$) of interest, of order $\E < \mathcal{O}(\whd^c)$ where $c \in (0, 2)$. If we expect $a \simeq b$ then $\E < \mathcal{O}(\whd^2)$. Therefore the difference between the two approximations will be small until we are close to $\whd = 1$, which is the resonant frequency peak. This supports the arguments made in connection with the results in~\cref{fig:AffinityPlot}.

\end{appendix}

\bibliography{BV_in_sim.bib}

\end{document}